\documentclass[twocolumn]{aastex701}
\received{March 13, 2026}
\revised{April 4, 2026}
\accepted{April 7, 2026}
\begin{document}

%% ======== Title & Authors ===================================================================================
%% ---------- Title -------------------
\title{Plasma Dynamics of Radiative Cooling Accretion Flow in AM Herculis with XRISM}
%% ---------- Authors -------------------
\author[orcid=0000-0002-2359-1857,sname='Terada']{Yukikatsu Terada}
%\author{Yukikatsu Terada}
\affiliation{Graduate School of Science and Engineering, Saitama University}
\affiliation{Institute of Space and Astronautical Science (ISAS),Japan Aerospace Exploration Agency (JAXA)}
\email[show]{terada@mail.saitama-u.ac.jp}  

\author[orcid=0000-0002-9709-5389,sname='Mori']{Kaya Mori} 
%\author{Kaya Mori} 
\affiliation{Columbia Astrophysics Laboratory, Columbia University}
\email{kaya@astro.columbia.edu}

\author[orcid=0000-0001-6665-2499,sname='Hayashi']{Takayuki Hayashi} 
%\author{Takayuki Hayashi} 
\affiliation{Department of Physics, Graduate School of Science, Kyoto University}
\email{hayashi.takayuki.3f@kyoto-u.ac.jp}

\author[orcid=0000-0002-6653-4975,sname='Bridges']{Gabriel L. Bridges} 
%\author{Gabriel L. Bridges} 
\affiliation{Columbia Astrophysics Laboratory, Columbia University}
\email{glb2139@columbia.edu}

\author[sname='Ishida']{Manabu Ishida} 
\affiliation{Institute of Space and Astronautical Science (ISAS),Japan Aerospace Exploration Agency (JAXA)}
\email{ishida@astro.isas.jaxa.jp}

\author[orcid=0000-0003-3441-9355,sname='Schwope']{Axel D. Schwope} 
%\author{Axel D. Schwope} 
\affiliation{Leibniz-Institut fur Astrophysik Potsdam}
\email{aschwope@aip.de}

\author[orcid=0009-0002-1729-8416,sname='Kimura']{Mariko Kimura} 
\affiliation{College of Science and Engineering, Kanazawa University}
\email{mariko-kimura@se.kanazawa-u.ac.jp}

\author[orcid=0000-0003-1130-5363,sname='Nobukawa']{Masayoshi Nobukawa} 
%\author{Masayoshi Nobukawa} 
\affiliation{Nara University of Education}
\email{nobukawa@cc.nara-edu.ac.jp}

\author[orcid=0000-0002-7004-9956,sname='Buckley']{David A. H. Buckley} 
%\author{David A. H. Buckley} 
\affiliation{South African Astronomical Observatory}
\affiliation{Department of Physics, University of the Free State}
\affiliation{Department of Astronomy, University of Cape Town}
\email{dah.buckley@saao.nrf.ac.za}

\author[orcid=0000-0001-6135-1144,sname='Balman']{\c{S}\"olen Balman} 
%\author{\c{S}\"olen Balman} 
\affiliation{Faculty of Science, Istanbul University}
\affiliation{Faculty of Engineering and Natural Sciences, Kadir Has University}
\email{solen.balman@istanbul.edu.tr}

\author[sname='Ichikawa']{Taichi Ichikawa} 
\affiliation{Graduate School of Science and Engineering, Saitama University}
\email{t.ichikawa.534@ms.saitama-u.ac.jp}

\author[sname='Matsumura']{Atsuto Matsumura} 
\affiliation{Science, Tokyo Metropolitan University}
\affiliation{Institute of Space and Astronautical Science (ISAS),Japan Aerospace Exploration Agency (JAXA)}
\email{matsumura-atsuto@ed.tmu.ac.jp}

\author[orcid=0009-0002-4998-9879,sname='']{Mai Takeo} 
%\author{Mai Takeo} 
\affiliation{Graduate School of Science and Engineering, Toyama University}
\email{takeo@sci.u-toyama.ac.jp}

\author[orcid=0000-0002-3681-145X,sname='']{Charles J. Hailey} 
\affiliation{Columbia Astrophysics Laboratory, Columbia University}
\email{chuckh@astro.columbia.edu}

\author[orcid=0000-0001-8722-9710,sname='Ramsay']{Gavin Ramsay} 
%\author{Gavin Ramsay} 
\affiliation{Armagh Observatory and Planetarium}
\email{gavin.ramsay@armagh.ac.uk}

\author[sname='']{Antonio Rodriguez} 
\affiliation{Department of Astronomy, California Institute of Technology}
\email{acrodrig@astro.caltech.edu}

\author[orcid=0009-0002-0326-3378,sname='Walker']{Samantha Walker} 
\affiliation{Department of Physics and Astronomy, Barnard College, Columbia University}
\email{sfw2117@barnard.edu}

%% ======== Abstract ==============================================================================
\begin{abstract}
%%%%%%%%%% Limit 250 words %%%%%%%%%%%%
We present XRISM/Resolve high-resolution X-ray spectroscopy of the prototypical magnetic cataclysmic variable AM Herculis. 
All satellite lines of highly ionized Fe are fully resolved. Lighter-element lines (Si, S, Ca) show 2–3 eV widths 
consistent with purely thermal broadening, while the broader 6–7 eV Fe lines require additional bulk Doppler broadening.
Spin-phase–resolved modulations are clearly detected in the Fe XXV and Fe XXVI lines, 
with semi-amplitudes of $81.8\pm6$ km s$^{-1}$ and $132.5\pm9$ km s$^{-1}$, 
and mean velocities of $143.6\pm6$ km s$^{-1}$ and $225.6\pm8$ km s$^{-1}$, respectively.
After removing these bulk Doppler shifts, we obtain intrinsic Doppler widths of $5.23_{-0.15}^{+0.16}$ eV for Fe XXV and $6.23_{-0.18}^{+0.19}$ eV for Fe XXVI, directly revealing gradients of bulk velocity and temperature in the cooling-flow plasma.
We additionally examined the resonance anisotropy predicted by Terada et al. (1999, 2001): the equivalent widths of the Fe XXV and Fe XXVI resonance lines increase at the pole-on phase by factors of 1.30–1.35, in positive correlation with their oscillator strengths. Combining
 XRISM with simultaneous NuSTAR data and PSAC/MCVSPEC plasma models, we derive a self-consistent shock temperature of $24.0\pm0.1$ keV and shock velocity of $1{,}116\pm2$ km s$^{-1}$.
Radiative-transfer simulations of the resonance lines further constrain the shock density to $\approx(5\text{–}6)\times10^{15}$ cm$^{-3}$, providing a new density diagnostic for accretion columns.
The resulting accretion-column geometry has a height of $200\text{–}300$ km and a radius of $200\text{–}400$ km.
\end{abstract}

%% Keywords ---------------------------------------------------------------------------------------
%% The AAS Journals now uses Unified Astronomy Thesaurus (UAT) concepts: https://astrothesaurus.org

%% 1. Introduction ===============================================================================================
\section{Introduction}\label{section:Introduction} % Added comment to avoid empty anchor warning%% ===============================================================================================================

%%----------------------------
%%% MCV introduction x
%%----------------------------
Magnetic Cataclysmic Variables (MCVs) are close binary systems consisting of a Roche\mbox{-}lobe filling late-type star and a magnetized white dwarf (WD). They are commonly grouped into intermediate polars (IPs) and polars. 
IPs host WDs that rotate asynchronously with the binary orbital period and have magnetic field strengths of $B\sim0.1\rm{-}10$~MG, strong enough to truncate the inner accretion disk. 
Polars, in contrast, contain WDs with stronger fields, typically $B\sim10\rm{-}240$~MG, and rotate mostly synchronously with the binary period. 
In both classes, plasma accreting onto the WD magnetic poles becomes supersonic and releases gravitational potential energy through a standing shock near the WD surface. 
The post-shock plasma is initially heated to a high temperature related to the depth of the gravitational potential of the WD and cools via thermal bremsstrahlung and cyclotron emission as it falls onto the surface. 
This process produces a magnetically confined, vertically stratified accretion column with gradients in temperature and density between the shock and the WD surface, which emits thermal X-rays with $kT \sim 10\rm{-}50$ keV \citep{1990SSRv...54..195C}.

%%----------------------------
%%% Importance of MCV spectral study
%%----------------------------
MCVs represent the most abundant class of compact binaries and hard X-ray sources ($E > 10$~keV) in our galaxy. 
In the solar neighborhood, the first volume-limited sampling of CVs within 150~pc found that about one third of them are magnetic systems \citep{Pala2020}, and numerous MCVs have been discovered by all-sky X-ray surveys  \citep{Schwope2025}. 
The majority of the two thousand Chandra X-ray point sources in the Galactic Center are likely MCVs \citep{Muno2009}. 
Diffuse X-ray emission in the Galactic Center and Ridge is believed to contain a large number of unresolved MCVs \citep{2008A&A...489.1121R, 2012ApJ...753..129Y, Perez2015, 2018PASJ...70R...1K}.
In order to disentangle the origin and composition of these diffuse X-ray emissions, it is crucial to characterize the X-ray spectral properties of individual MCVs and link them to their WD masses, magnetic fields, and accretion rates \citep{2006A&A...452..169R,2016ApJ...818..136X,2024ApJ...961..205K}. 
These efforts can be best performed by modeling high-resolution Fe lines and broadband X-ray spectra of MCVs. 

%%===========================================================
%% Figure 1. (fig.1, Catoon)
%%===========================================================
\begin{figure}[htb]
    \centering
    \includegraphics[width=0.85\linewidth]{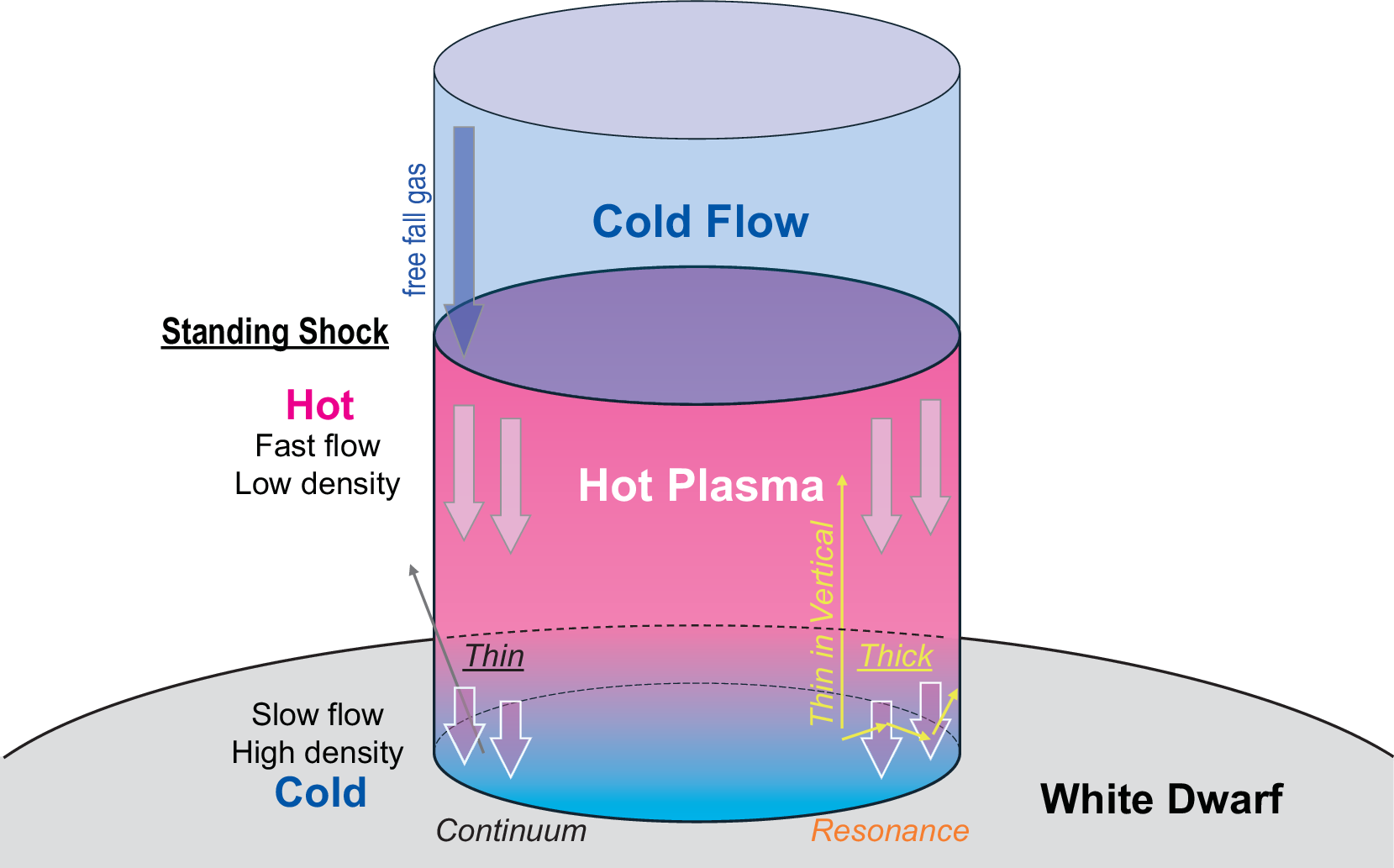}    
    \caption{A schematic illustration of the hot plasma in an accretion column on a magnetized white dwarf of an MCV is shown.
    The multi-temperature plasma structure and the mechanism of resonance anisotropy are also represented.}
    \label{fig:accretion_column_schematic}
\end{figure}
    
%%----------------------------
%%% CIE plasma 
%%----------------------------
The shock-heated plasmas in MCV rapidly reach collisional ionization equilibrium (CIE) because of their high number densities , $\sim 10^{13-17}$ cm$^{-3}$, which correspond to short radiative cooling times of $\sim$ s. 
As a result, MCVs are excellent astrophysical laboratories for studying CIE plasmas and associated processes within practical X-ray exposure times. 
This contrasts with CIE plasmas in galaxy clusters, which have much lower densities ($10^{0-5}$ cm$^{-3}$) and very long dynamical timescales ($\sim 10^{13-17}$ s), as well as non-equilibrium ionization (NEI) plasmas in supernova remnants (e.g, \citealt{1984Ap&SS..98..367M,1989A&A...221..385B,1990PASJ...42..279H} etc) and non magnetic CVs (\citealt{2025Univ...11..105B} and references therein). 
For example, a key open question in MHD plasma physics is how electrons and ions are heated to different temperatures across the standing shock.
To address this, several laser-bombardment experiments have attempted to reproduce a scaled-down post-shock region \citep{VBS2018}, 
but these laboratory studies are restricted to weak magnetic-field conditions where thermal bremsstrahlung cooling is dominant \citep{VBS2018}.  

%%----------------------------
%%% Plasma structure of MCVs / previous X-ray observations
%%----------------------------
In the standard picture originally proposed in 1973 \citep{1973PThPh..49..776H,1973PThPh..49.1184A}, the accretion column in MCVs is expected to show vertical gradients in plasma temperature, density, and velocity due to radiative cooling: as the gas falls toward the WD surface and its bulk velocity decreases, its temperature falls while its density rises, as illustrated in Figure \ref{fig:accretion_column_schematic}.  
This multi-temperature accretion-column scenario and its variants successfully accounted for the X-ray spectra of MCVs observed with Ginga, ASCA, and RXTE in the 1990s \citep{1995MNRAS.276..483D,1997ApJ...474..774F}.  
Over the past two decades, two main methods have been employed: either fitting broadband X-ray continuum spectra or Fe emission lines, both aimed primarily at constraining WD masses in MCVs. 
In the former, Suzaku and NuSTAR observations of IPs provided hard X-ray coverage above 10 keV, enabling robust measurements of shock temperatures and thus WD masses \citep{2010A&A...520A..25Y,Hailey2016, Suleimanov2019, Shaw2020}.  
In the latter, Fe-line diagnostics were used to derive ionization temperatures and infer WD masses \citep{Ezuka1999, Xu2016}, based on line intensities from neutral, He-like, and H-like Fe that remained only partially resolved with the CCD spectrometers on ASCA, Chandra, and XMM-Newton.  
These methods have yielded different WD masses that conflict with independent estimates obtained at other wavelengths \citep{AH2023, AH2024}.

While X-ray emission from the accretion column of MCVs is well modeled, its realization and detailed application to X-ray spectra have been limited due to the insufficient energy resolution and bandwidth of earlier X-ray spectroscopy data. 
For instance, the most recent X-ray spectral models have incorporated all essential physical and geometrical effects -- such as atomic lines, cyclotron cooling, separate electron and ion flow solutions, specific accretion rates, gravitational acceleration, and dipole magnetic field geometry \citep{2014MNRAS.441.3718H, Filor2025}. 
For given WD parameters and accretion rates, these models can readily produce temperature, density, and bulk velocity profiles in the accretion column and fit X-ray spectral data, as demonstrated for several IPs and polars. 
However, more refined X-ray spectral features arising from gas dynamics, such as gradients in bulk velocity and plasma density, have not been observable due to the lack of high-resolution X-ray spectral data. 

%%----------------------------
%%% XRISM
%%----------------------------
In 2023, the X-Ray Imaging and Spectroscopy Mission (XRISM) \citep{2025PASJ...77S...1T} was launched and carried out non-dispersive high-resolution X-ray spectroscopy with an energy resolution of 5 eV in the $1.7\rm{-}10$ keV band with a microcalorimeter called Resolve \citep{2025JATIS_Kelley_Resolve,2025JATIS_Ishisaki_Resolve}.
XRISM has effectively demonstrated its ability to quantitatively reveal the dynamics of gas turbulence within galaxy clusters and the kinematics of clumpy absorbers surrounding black holes in active galactic nuclei \citep{2025Natur.638..365X,2025Natur.641.1132X,2025Natur.646...57X,2024ApJ...977L..34X,2025ApJ...982L...5X,2025ApJ...985L..20X,2025ApJ...988L..58X,2025ApJ...993L..11X,2025PASJ..tmp..116A}, 
%\citep{2025Natur.638..365X,2025Natur.641.1132X,2025Natur.646...57X,2025ApJ...988L..58X}, 
as well as detailed diagnostics of plasma in supernovae, supernova remnants, and the Galactic Center region 
\citep{2024ApJ...973L..25X,2024PASJ...76.1186X,2025PASJ...77L...1X,2025PASJ...77S.193X,2025PASJ...77S.242X,2025A&A...702A.147X}.
%\citep{2024ApJ...973L..25X,2024PASJ...76.1186X,2025PASJ...77S.193X}.
Similarly, high-resolution XRISM spectroscopy data of MCVs are expected to enable more detailed diagnostics of MCVs regarding plasma density, temperature, and flow dynamics from different viewing angles over the spin phase.

%%----------------------------
%%% Paper structure
%%----------------------------
This paper presents the first results on magnetic cataclysmic variables obtained with XRISM/Resolve.  
The manuscript is organized as follows; section \ref{section:observation_reduction} describes the XRISM and NuSTAR X-ray observations of the MCV AM Herculis and summarizes the data reduction procedures.  
Section \ref{section:results} details the analysis of the XRISM/Resolve X-ray line spectra, revealing plasma dynamics by line Doppler effects of the spectral lines.
In Section \ref{section:discussion}, we interpret the observational findings using a plasma emission model together with the simultaneous NuSTAR spectrum and perform the radiative transfer simulations to determine the plasma properties of the accretion column.  
Finally, Section \ref{section:conclusion} summarizes the main results of this work.

%% 2. Observation & Data Reduction ===============================================================================
\section{Observation and Data Reduction}
\label{section:observation_reduction}
%% ===============================================================================================================

We performed a Target of Opportunity observation of the prototype and brightest polar, AM Herculis, with XRISM.
Since AM Herculis shows both high and low accretion states, we requested the American Association of Variable Star Observers (AAVSO) to begin monitoring its optical magnitude soon after the first cycle of XRISM's guest observation program starts in September 2024. 
The AAVSO monitoring campaign started on 8 October 2024, and we found that the source had entered a high state with a visual magnitude $V\sim13$ mag on 11 October 2024, immediately following a two-month low state at $V\sim15.3$ mag.
Thereafter, an X-ray ToO observation with the Swift satellite on 19 October 2024 confirmed that the 2--10 keV X-ray flux exceeded XRISM's ToO trigger threshold of $1\times10^{-10}$ erg s$^{-1}$ cm$^{-2}$. 
Then, the XRISM observation of AM Herculis (PI: Y.\ Terada and K.\ Mori) was carried out from 23:31:00 UTC on 21 October 2024 to 00:30:00 UTC on 26 October 2024, with a net exposure for Resolve of 184 ks. 
The observation identification number is OBSID = 201035010.  
The telescope was pointed at (RA, Dec)$_{\rm J2000.0}$ = (274.0554$^\circ$, 49.8678$^\circ$) with a roll angle of 243.2537 degrees.  
XRISM/Resolve was operated with no filter but with the gate valve (GV) closed, restricting its energy bandpass to 1.7--10 keV.  
XRISM/Xtend operated in nominal mode with the 1/8 window configuration.
In addition, we carried out a simultaneous observation of AM Herculis with NuSTAR \citep{2013ApJ...770..103H} from 07:11:08 UTC on 24 October 2024 to 14:41:09 UTC on 25 October 2024, with a net exposure of 62 ks (OBSID = 80960303002). 

\begin{figure}[thb]
    \centering
    \includegraphics[width=0.90\linewidth]{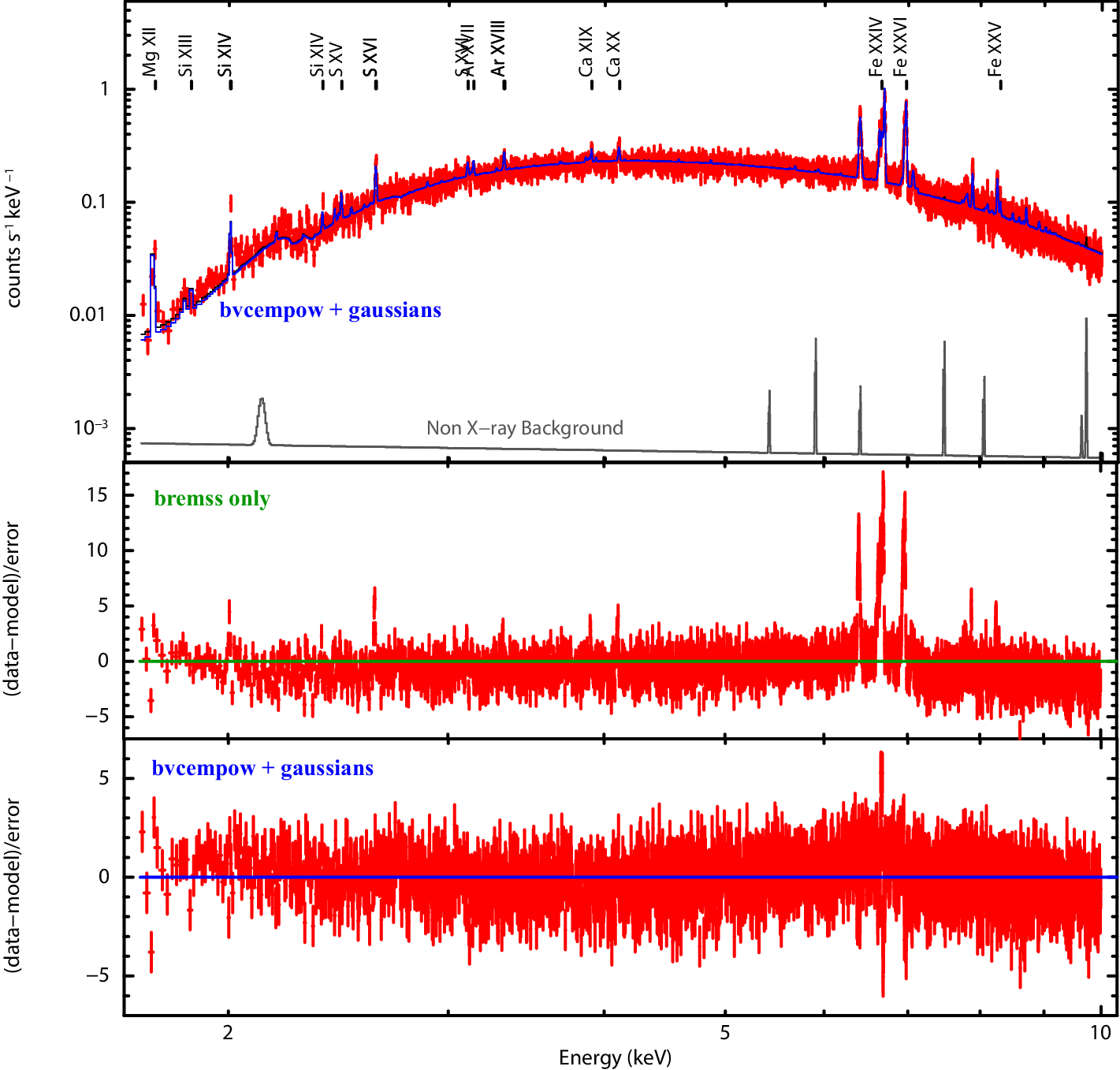}
    \caption{The phase-averaged X-ray spectrum of AM Herculis observed with XRISM/Resolve is plotted as red crosses. 
    In the upper panel, the best-fit 'bvcempow' model with three Gaussian components and the non X-ray background model (see the text) are indicated by blue and gray lines, respectively. 
    The middle and bottom panels show the residuals with respect to the 'bremsstrahlung' model and the 'bvcempow' model with three Gaussian components, respectively.}
    \label{fig:methods:phase_av_spec}
\end{figure}

The XRISM observation data were processed with the standard pre-pipeline and pipeline processes \citep{2021JATIS...7c7001T,2025JATIS_Hayashi_SOT}, with TLM2FITS version '005\_002.20Jun2024\_Build8.012' and PROCVER '03.00.013.009', respectively.
For the analysis, we used the XRISM ftools in the HEAsoft package (version 6.36) together with the CALDB XRISM version 12.
Resolve events were extracted from the standard cleaned event file available in the data archive.
Following the instructions in the XRISM quick start guide (version 2.3, dated 18 September 2024), we applied additional screening based on the PI, RISE\_TIME, and DERIV\_MAX parameters as follows: {\tt (PI$\geq$600) \&\& ((RISE\_TIME+0.00075$\times$DERIV\_MAX)$>$46) \&\& ((RISE\_TIME+0.00075$\times$DERIV\_MAX)$<$58)}.
In the following analysis of XRISM/Resolve, we used high primary grade (Hp) \citep{2025JATIS_Ishisaki_Resolve}.
For spectral fitting, we used XSPEC version 12.15.1 in the HEAsoft package and AtomDB version 3.1.3 for the {\it apec} family for XSPEC \citep{2001ApJ...556L..91S}. 
The non X-ray background (NXB) spectrum for XRISM is provided by the XRISM collaboration; we used the provisional version 2 model for this observation.
The source signal exceeds the NXB by one to two orders of magnitude, as illustrated in Figure \ref{fig:methods:phase_av_spec}.

We processed NuSTAR data using {\tt nupipeline} in the NuSTAR Data Analysis Software (NuSTARDAS version 2.1.2). 
There are two flight module data points represented by FPMA and FPMB. We extracted NuSTAR FPMA and FPMB spectra from a $r = 60''$ circular region around the source centroid in the cleaned event files. 
The background spectra were extracted from a source-free circular region of $r = 200''$. The source was detected above the background up to 60 keV, for a total of $1.9\times10^5$ net counts combined between FPMA and FPMB data.

%% 3. Results ====================================================================================================
\section{Observational Results}
\label{section:results}
%% ===============================================================================================================

%%===========================================================
%% Figure 2 (fig 7), Phase averaged specrum  --- section 3
%%===========================================================
%%===========================================================

%-----------------------------------------------------------------------------------------------------------------
\subsection{Phase-averaged Spectrum with Resolve}
\label{section:results:phase_averaged_overview}
%-----------------------------------------------------------------------------------------------------------------

\begin{figure*}[ht]
    \centering
    \includegraphics[width=0.95\linewidth]{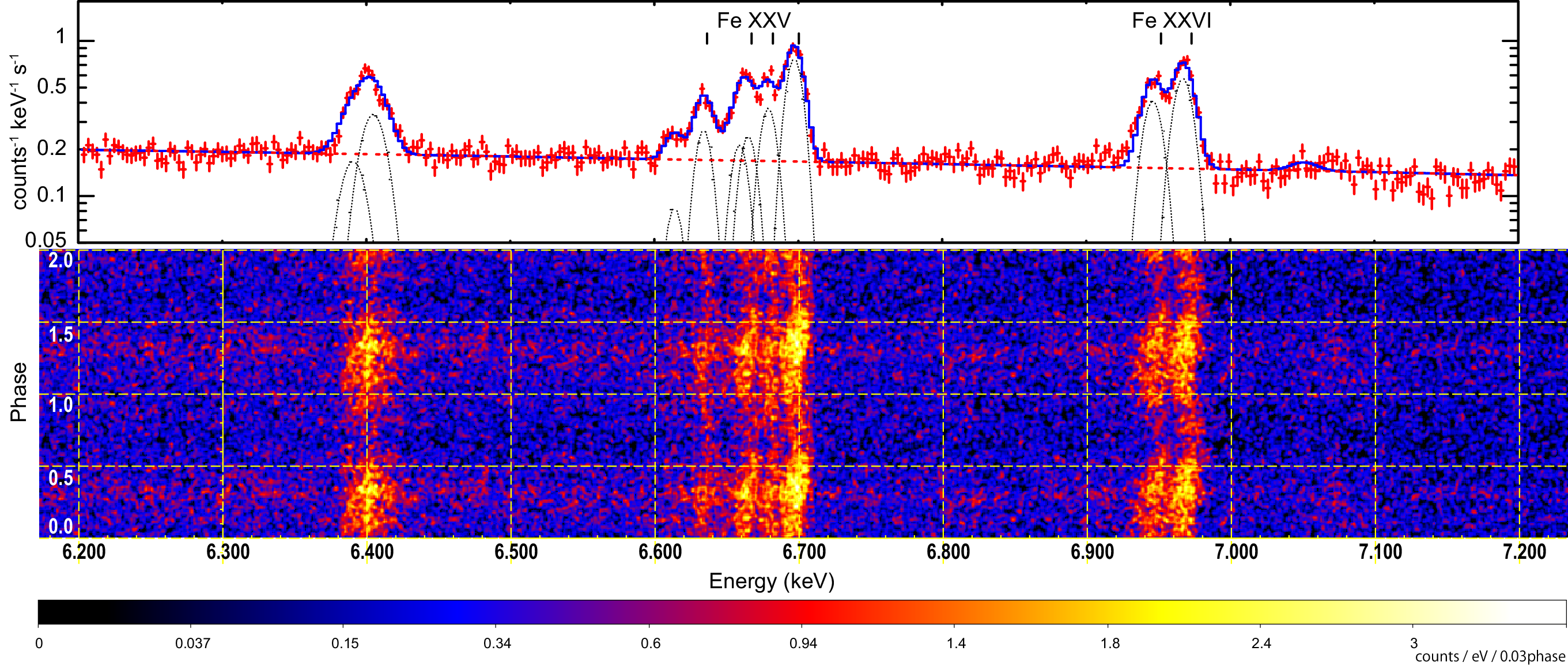}
    \caption{(Upper) The phase-averaged X-ray spectrum of AM Herculis with XRISM, focusing on the Fe-K lines, is shown in red crosses. 
    The optimal Gaussian models representing Fe-K fluorescence, Fe XXV, and Fe XXVI lines are illustrated with black dotted lines.
    (Lower) A two-dimensional plot of photon energy versus spin phases is presented, computed using the ephemeris from equation (1) in \citet{2020A&A...642A.134S}. 
    To enhance the visibility of repeating features, the data is displayed over two cycles.
    The lower panel represents the scale of the two-dimensional plot.}
    \label{fig:spec_energy_phase_fe}
\end{figure*}

%%% ------------------------------------
%%% phase averaged, single bremss.
%%% ------------------------------------
The phase-averaged X-ray spectrum of AM Herculis obtained with XRISM is shown in Figure \ref{fig:methods:phase_av_spec}. 
We expect X-ray emission from a multi-temperature plasma, but within the limited bandpass of GV closed conditions in the 1.7 -- 10 keV range, the continuum spectrum is roughly reproduced by a single bremsstrahlung model ("bremss" in XSPEC) with a temperature of $33.4_{-1.4}^{+1.5}$ keV and an emission measure (EM) of $(6.55\pm0.03)\times10^{54}$ cm$^{-3}$ with C-statistics of 19484.1 for degrees of freedom (d.o.f) of 2697.
The observed X-ray flux in the 2–10 keV band is $8.4\times 10^{-11}$ erg cm$^{-2}$ s$^{-1}$.
As shown in Figure \ref{fig:methods:phase_av_spec} (middle), the high energy resolution spectroscopy of Resolve enables us to separate the atomic lines of Si, S, Ar, and Fe from the continuum emission.
Ions identified in the spectrum based on the AtomDB database are listed in Table \ref{tab:lines_identified}.
In the complex of Fe lines at energy $E \sim 6\rm{-}7$ keV, we clearly detected the resonance, intercombination, and forbidden satellite lines from Fe XXV (He-like) and Fe XXVI (H-like), as shown in Figure \ref{fig:spec_energy_phase_fe} (top); the Chandra HETG marginally resolved these features in earlier observations \citep{2007ApJ...658..525G}, but XRISM provides, for the first time, a distinctly clear separation of these satellite lines.

%%% ------------------------------------
%%% phase averaged, multi-kT
%%% ------------------------------------
To reproduce the overall spectrum, including the atomic lines in the 1.7--10 keV band with continuum emission, we employed the multi-temperature CIE plasma model ‘bvcempow’ in XSPEC, combined with photoelectric absorption (‘TBabs’) and reflection (‘reflect’) components, plus three Gaussian profiles for the Fe fluorescent lines (i.e., two Fe K$_\alpha$ and K$_\beta$ lines).
The best-fit model is plotted as the blue lines in Figure \ref{fig:methods:phase_av_spec}, and the corresponding best-fit parameters are listed in Table \ref{tab:methods:wide_band_fit} under the label ‘wide-band’.
Here, $\alpha$ and $kT_{\rm max}$ for ‘bvcempow’ represent the index of the power-law emissivity function and the maximum temperature, respectively, where the EM satisfies the relation $d\rm{EM}=(kT/kT_{\rm max})^{\alpha-1}\, dT/T_{\rm max}$.  
The derived $kT_{\rm max}$ values are slightly higher than the temperature obtained from a single-temperature bremsstrahlung fit (33.4 keV), which corresponds to the EM–weighted mean temperature, including contributions from cooler components.

%--------------------------------------------------------------------------------
% Table 1. (new table) LInes identified 
%--------------------------------------------------------------------------------
\begin{deluxetable*}{cl}
\digitalasset
\tablewidth{0pt}
\tablecaption{Atomic lines in the phase-averaged spectrum of AM Herculis}
\tablehead{
\colhead{Ion$^\dagger$}                 & \colhead{Line energies (keV)} 
}
\startdata
    Si XIV   & 2.004, 2.006, 2.377 \\
    S XV     & 2.430, 2.461 \\
    S XVI    & 2.623, 3.107 \\
    Ar XVII  & 3.104, 3.140 \\
    Ar XVIII & 3.323\\
    Ca XIX   & 3.902\\
    Fe XXIV  & 6.662, 6.645$^\ddagger$, 6.655$^\ddagger$ \\
    Fe XXV   & 6.637, 6.668, 6.682, 6.700, 7.882, 8.295\\
    Fe XXVI  & 6.952, 6.973, 8.246, 8.253 
\enddata
\tablecomments{$\dagger$ Ions identified using AtomDB version 3.1.3, with larger line emissivities than $10^{-18}$ ph cm$^3$ s$^{-1}$. $\ddagger$ Dielectronic recombination satellite Lines.}
\label{tab:lines_identified}
\end{deluxetable*}
%-----------------------------------------------------------------------------------------------------------------
\subsection{Line Shift and Width in Phase-averaged Spectrum}
\label{section:results:phase_averaged_lines}
%-----------------------------------------------------------------------------------------------------------------

%%===========================================================
%% Figure 3. (fig.2, phase average spectrum, Energy vs Phase 2D plot)
%%===========================================================
%%===========================================================

%--------------------------------------------------------------------------------
% Table 2. (table 1) phase averaged models
%--------------------------------------------------------------------------------
\begin{deluxetable*}{lccccc}
\digitalasset
\tablewidth{0pt}
\tablecaption{Best-fitting parameters for the phase-averaged spectrum of AM Herculis$^\dagger$}
\tablehead{
\colhead{Parameter}                 & \colhead{wide-band} &  \colhead{Si}       &   \colhead{S}         &   \colhead{Ca}        & \colhead{Fe} 
}
\startdata
    Energy band (keV)               & 1.7 - 10         & 1.75 - 2.2             & 2.40 - 2.75           & 3.5 - 4.5             & 6.2 - 7.2 \\
%   \hline %------------------------------------------------------------------------------------------------------------------------------------
    nH ($10^{22}$ cm$^{-2}$)        & $1.69\pm0.06$    & 1.69 fix               & 1.69 fix              & 1.69 fix              & 1.69 fix \\   
    $\alpha$                        & $0.40\pm0.04$    & $1.46_{-0.13}^{+0.19}$ & $0.93_{-0.16}^{+0.38}$& $0.65_{-0.23}^{+0.11}$& $0.28\pm0.03$\\
    $kT_{\rm max}$                  & $37\pm2$         & 37 fix                 & 37 fix                & 37 fix                & 37 fix\\
    'bvcempow' Doppler velocity (km s$^{-1}$)  
                                    & $223\pm8$        & $17_{-5}^{+70}$        & $137\pm44$            & $113_{-46}^{+49}$     & $227\pm9$\\
    'bvcempow' Doppler width (km s$^{-1}$)     
                                    & $289_{-9}^{+10}$ & 0.0                    & $<221$                & $185_{-93}^{+83}$     & $289_{-10}^{+11}$\\
    \hline %--------------------------------------------------------------------------------------------------------------
    Gaussian$^\ddagger$ Doppler velocity (km s$^{-1}$) 
                                    & $26\pm20$       & N/A                     & N/A                   & N/A                   & 26 fix\\
    Gaussian$^\ddagger$ Doppler width (km s$^{-1}$)           
                                    & $501_{-23}^{+25}$& N/A             & N/A                   & N/A                   & 501 fix\\   
    \hline %--------------------------------------------------------------------------------------------------------------------------------
    C-statistics                    & 3827.93          & 139.2                  &  113.95               & 354.33                & 877.5\\
    d.o.f.                          & 2684             & 151                    &  115                  & 327                   & 402
\enddata
\tablecomments{$\dagger$ We applied "TBabs * reflect * bvcempow" with three Gaussian model.\\
$\ddagger$ Doppler velocity and line width for three Gaussian components, modeling the Fe K$_\alpha$ and K$_\beta$ lines, taken to be identical for all three lines. Please refer to the text for the corresponding definitions.}
\label{tab:methods:wide_band_fit}
\end{deluxetable*}

%%% ------------------------------------
%%% Doppler velocity and width
%%% ------------------------------------
Fitting the Resolve spectra in all bands with the CIE plasma model ‘bvcempow’ (Figure \ref{fig:methods:phase_av_spec} and Table \ref{tab:methods:wide_band_fit}) provides the Doppler shifts and line broadening, which are listed in the table as Doppler velocity and Doppler width in units of km s$^{-1}$. 
The Doppler velocity is derived from the difference between the observed line-center energy and the corresponding rest-frame laboratory energy for each ion species in the AtomDB database \citep{2001ApJ...556L..91S}, where a positive value indicates a redshift.
The Doppler width characterizes the intrinsic line width of the object (in the photon-space spectrum); the contribution from the detector energy resolution has already been removed in the fitting procedure (in the count-space spectrum), since it is encoded in the energy response function.
Because the line spread function and energy resolution in the response function of Resolve are accurately calibrated \citep{2025JATIS_Ishisaki_Resolve}, this instrument is capable of detecting line broadening at the sub-eV scale for narrow spectral lines, even when this intrinsic width is smaller than the nominal energy resolution of 5–7 eV.
Therefore, for the phase-resolved spectrum, we measure a redshifted Doppler velocity of about 220 km s$^{-1}$ and a positive line broadening of about 290 km s$^{-1}$ (Table \ref{tab:methods:wide_band_fit} under the ‘wide-band’ label).
We also detected broad Fe fluorescent lines with a width of about 500 km s$^{-1}$, significantly exceeding the rotational velocity at the surface of the WD.

%--------------------------------------------------------------------------------
% Table 3. (table 2) phase-averaged narrow band models
%--------------------------------------------------------------------------------
\begin{deluxetable*}{lcccccccc}
\digitalasset
\tablewidth{0pt}
\tablecaption{Best-fitting parameters in narrow-band phase-averaged spectra$^\dagger$}
\tablehead{
    \colhead{Parameter}            &  \colhead{Si XIV} & \colhead{S XVI}     & \colhead{Ca XX,XIV}   & \colhead{Fe XXV}      & \colhead{Fe XXVI}  
}
\startdata
    Energy band (keV)              & 1.9 - 2.1         & 2.40 - 2.75         & 3.5 - 4.5         & \multicolumn{2}{c}{6.2 - 7.2} \\
%   \hline %--------------------------------------------------------------------------------------------------------------------------------
    'bvcempow' Doppler velocity (km s$^{-1}$)  
                                   & $222_{-3}^{+348}$ & $163_{-289}^{+151}$ & $55_{-41}^{+43}$  & $126\pm10$  & $255\pm11$        \\
    'bvcempow' Doppler width (km s$^{-1}$)     
                                   & $<150$            & $243_{-85}^{+109}$  & $159_{-62}^{+81}$ & $253\pm8$   & $290_{-9}^{+10}$  \\
    \hline %--------------------------------------------------------------------------------------------------------------------------------
    Gaussian$^\ddagger$ Doppler velocity (km s$^{-1}$)
                                   & N/A               & N/A                 & N/A               & \multicolumn{2}{c}{$36\pm20$}\\
    Gaussian$^\ddagger$ Doppler width (km s$^{-1}$)
                                   & N/A               & N/A                 & N/A               & \multicolumn{2}{c}{$417\pm23$}\\    
    \hline %--------------------------------------------------------------------------------------------------------------------------------
    C-statistics                   & 58.38             & 84.36               &  349.52           & \multicolumn{2}{c}{426.95}       \\
    d.o.f.                         & 63                & 77                  &  325              & \multicolumn{2}{c}{321}           
\enddata
\tablecomments{$\dagger$ We applied "TBabs * reflect * bvcempow" with three Gaussian model.\\
$\ddagger$ Doppler velocity and line width for three Gaussian components, modeling the Fe K$_\alpha$ and K$_\beta$ lines, taken to be identical for all three lines. Please refer to the text for the corresponding definitions.}
 \label{tab:methods:line_fit}
\end{deluxetable*}

%%% ------------------------------------
%%% narrow band line fit
%%% ------------------------------------
The remaining residuals from the spectral model in the wide-band, phase-averaged spectral fit (Figure \ref{fig:methods:phase_av_spec}, bottom) reflect differences in Doppler velocities and line widths among the species.
Therefore, to determine the Doppler shift and width of each individual line, we fit the phase-averaged spectrum with the same model (“TBabs * reflect * bvcempow” plus three Gaussians), but over a narrower energy range, fixing the continuum parameters (namely, the temperature of “bvcempow" and the photoabsorption and reflection parameters).
Thanks to the precise calibration of the Resolve line spread function \citep{2025JATIS_Kelley_Resolve}, we measured the intrinsic line widths of $\Delta E = 2\rm{-}7$ eV, as listed in Table \ref{tab:methods:wide_band_fit}.
These widths are consistent with the combined effects of natural broadening ($\sim 1$ eV), thermal Doppler broadening ($\sim 2\rm{-}3$ eV), and bulk Doppler motions, resulting in line centroid shifts. 
The Si, S, and Ca lines are narrower, with widths of $\Delta E \sim 2$ eV, indicating negligible bulk Doppler effects, whereas the Fe lines are broader, at $\sim 7$ eV, reflecting a stronger contribution from bulk gas flow. 

%%% ------------------------------------
%%% Ions line fit
%%% ------------------------------------
In the procedure described above for fitting lines of each species, however, the fit to the Fe band still exhibit relatively large Cash statistics in relation to the degrees of freedom; i.e., the overall Fe-line profile cannot be well characterized by a single Doppler shift and width parameter.
This indicates that Fe XXV and Fe XXVI have different Doppler widths and shifts, implying that their ionization states must be treated separately.
We, therefore, carried out additional narrow-band fits for Si XIV, S XVI, Ca XIV/XX, Fe XXV, and Fe XXVI individually.
The results, summarized in Table \ref{tab:methods:line_fit}, confirm the same conclusion that light elements exhibit narrower widths of 2–3 eV, while the Fe lines show broader widths of 6–7 eV.
Because the thermal Doppler widths of these lines in the multi-temperature plasma are determined by the temperature of peak line emissivity (2–4 eV), the measured widths for the lighter elements are consistent with purely thermal broadening. 
In contrast, the larger widths of Fe XXV and Fe XXVI indicate an additional Doppler contribution from bulk gas motions.
The details will be analyzed in the following section.

%-----------------------------------------------------------------------------------------------------------------
\subsection{Spin-phase Fe-line modulations}
\label{section:results:phase_resolved}
%-----------------------------------------------------------------------------------------------------------------

%%===========================================================
%% Figure 4. (fig.8, energy vs phase 2D)
%%===========================================================
\begin{figure*}[hbt]
    \centering
    \includegraphics[width=0.95\linewidth]{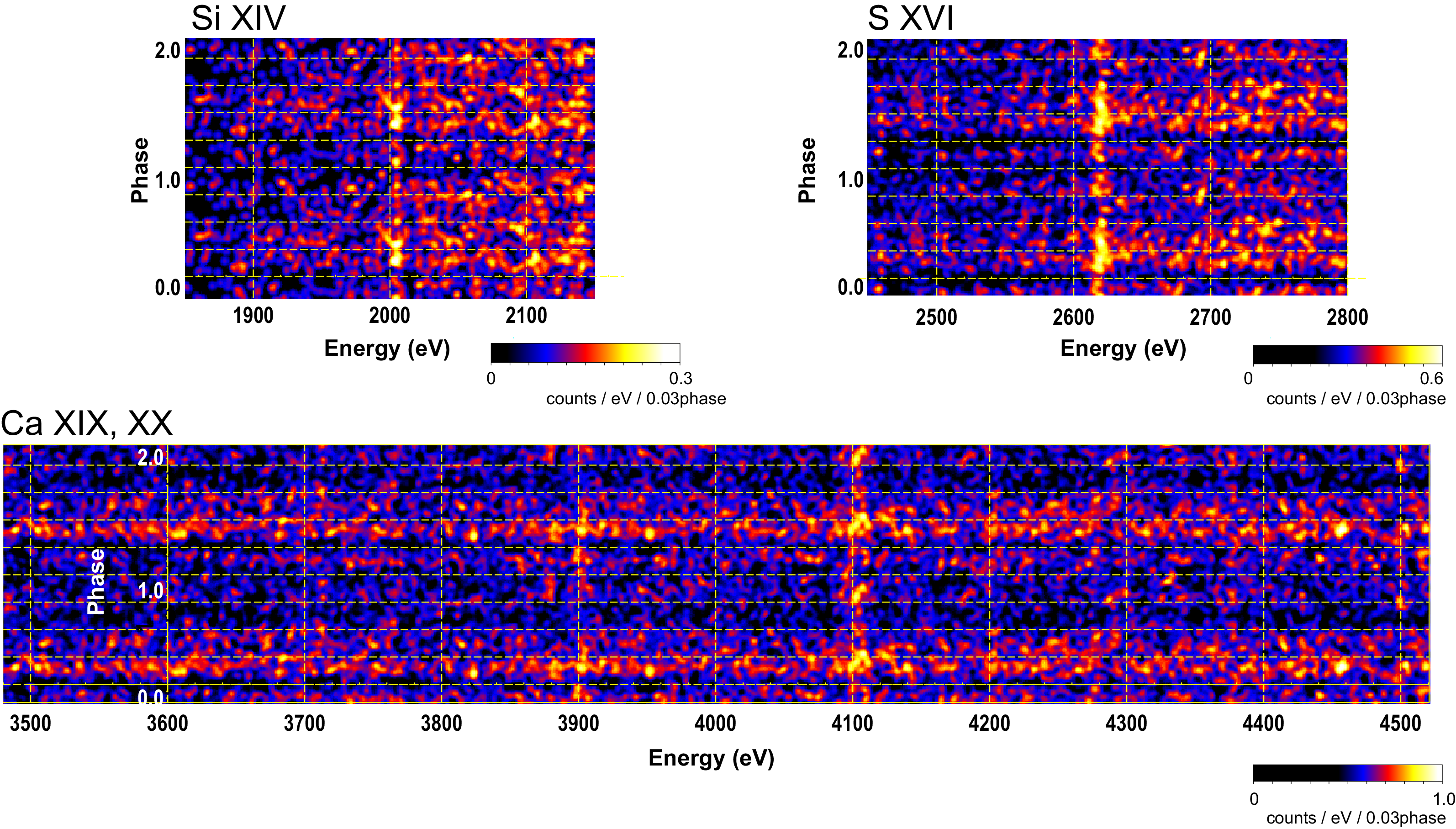}
    \caption{Same as Figure \ref{fig:spec_energy_phase_fe} (bottom), 
    but showing the energy bands around the Si XIV, S XVI, and Ca XIX–XX lines in the top-left, top-right, and bottom panels, respectively.}
    \label{fig:methods:spec_energy_phase_light}
\end{figure*}
%%===========================================================

%%%%%
To separate the bulk Doppler shifts (i.e., those arising from gas flow motions and changing with spin phase) from the Doppler broadening seen in the phase-averaged spectra, we performed a phase-resolved analysis of the X-ray spectrum, adopting the known orbital (spin) period of 3.1 hr.
We first applied a barycentric correction to the arrival times of the Resolve cleaned events at the target position using the `barycen` tool in the HEAsoft package \citep{2025JATIS_Terada_Time}.
Next, we calculated the spin-orbital phase based on the ephemeris given in equation (1) of \citet{2020A&A...642A.134S}, where phase zero corresponds to the inferior conjunction of the secondary star.
Note that the energy axis is not corrected for Doppler shifts due to the binary motion in these plots.
Consequently, the XRISM/Resolve spectra, folded by spin phases derived from the ephemeris of \citet{2020A&A...642A.134S}, clearly show modulations attributable to bulk Doppler shift, as illustrated in Figure \ref{fig:spec_energy_phase_fe} (bottom).
While such modulation of Fe XXVI line energies was previously reported with Chandra HETG \citep{2007ApJ...658..525G} and XMM-Newton EPIC-pn \citep{2020A&A...642A.134S} at lower significance, XRISM provides the first unambiguous detection of spin modulations in all satellite lines of Fe XXV and Fe XXVI arising from shock-heated plasma, as well as the Fe fluorescent K line.

%%%%
We further examined the spin-orbital modulations for lighter elements, as shown in Figure \ref{fig:methods:spec_energy_phase_light}.
As a result, spin-modulated features appear only in the Fe lines, while no such modulation is detected for the lighter elements.
This result is consistent with the phase-averaged Doppler analyzes, which do not require an additional bulk Doppler component for the lighter elements.

\begin{figure*}[hbt]
    \centering
    \includegraphics[width=0.95\linewidth]{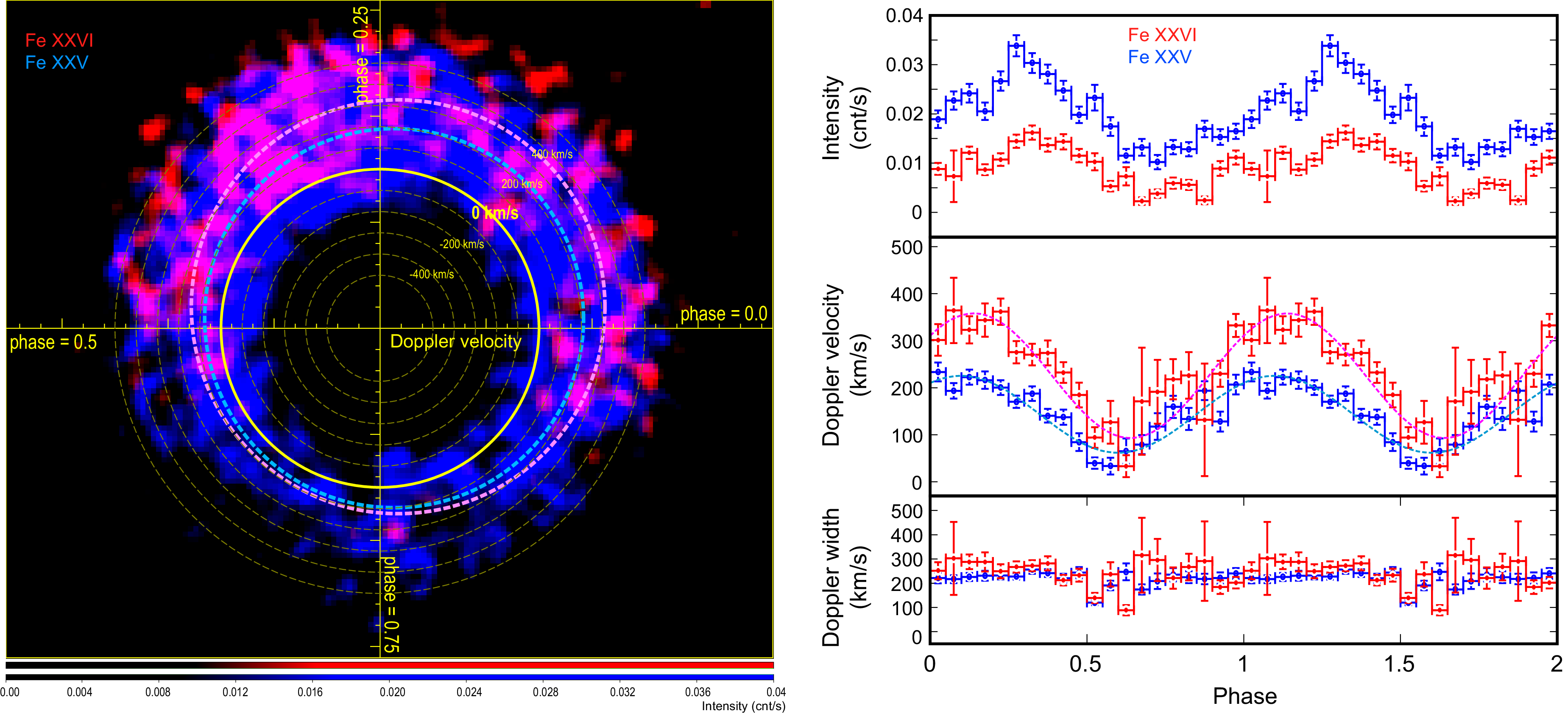}
    \caption{(Left) The Doppler velocities of highly-ionized Fe lines are displayed in polar coordinates as a function of phase, with the Fe XXV lines in blue and the Fe XXVI lines in red. 
    The spin-orbital phase is calculated using the ephemeris given by equation (1) in \citet{2020A&A...642A.134S}, with phase zero defined at the inferior conjunction of the secondary star.
    The radial and angular dimensions represent the Doppler velocity and the phase, respectively. 
    Velocity coordinates span from -500 km s$^{-1}$ to 500 km s$^{-1}$, while phase coordinates range from 0.0 to 1.0, highlighted in yellow. 
    (Right) From top to bottom, the panels display the line intensities, Doppler velocities, and Doppler widths of the Fe XXV and Fe XXVI lines, shown in blue and red, respectively, as a function of phase. 
    The errors are presented at a 1.0 $\sigma$ confidence level.
    In the left panel and in the second panel on the right, the best-fitting sinusoidal curves for the Doppler velocities of the Fe XXV and Fe XXVI lines are displayed as dashed magenta and cyan lines, respectively.}
    \label{fig:velocity2d_doppler}
\end{figure*}

%%%%
To quantitatively analyze the modulation of the line energies of Fe XXV and Fe XXVI, we derived the bulk Doppler-shift velocities from the differences between the photon energies measured with Resolve and the corresponding theoretical line energies from AtomDB v3.1.3 \citep{2001ApJ...556L..91S} (Table \ref{tab:lines_identified}).
This calculation was first carried out for individual lines (Fe XXV resonance, intercombination 1 and 2, forbidden; Fe XXVI resonance 1 and 2) and then combined by ion species (Fe XXV and Fe XXVI).
Because these lines are well separated in the phase-resolved Fe spectra, and we adopt narrow energy bands when calculating Doppler velocities, contamination from neighboring lines in the measured Doppler velocities is negligible.
The resulting velocities as a function of phase are shown in polar coordinates in Figure \ref{fig:velocity2d_doppler} (left). 
In this representation, the spectral continuum component has been subtracted, and the result is shown in units of counts s$^{-1}$ after normalization by the polar-coordinate image corresponding to a flat (phase- and velocity-independent) distribution.
In this diagram, the points outside the yellow circle marked as 0 km s$^{-1}$ indicate that both lines are redshifted at all phases. 
Moreover, the Fe XXVI data are located at larger radii than the Fe XXV data, implying that Fe XXVI in the hotter upper layers of the accretion column has a greater bulk Doppler velocity compared to Fe XXV in the cooler, lower layers.

%%===========================================================
%% Figure 5. (fig.3, Velocity vs Phase 2D plot, Doppler vs phase)
%%===========================================================
%%===========================================================

Numerically, Figure \ref{fig:velocity2d_doppler} (right) is produced by fitting Gaussians in velocity space across phase.  
We then modeled the spin modulation of bulk\mbox{-}Doppler velocities of Fe XXV and Fe XXVI with a sinusoidal function.
The best-fit curves, shown as dashed cyan and magenta lines in the second panel of Figure \ref{fig:velocity2d_doppler} (right), yield semi-amplitudes of $81.8\pm6$ km s$^{-1}$ and $132.5\pm9$ km s$^{-1}$, and mean velocities of $143.6\pm6$ km s$^{-1}$ and $225.6\pm8$ km s$^{-1}$, respectively.
These values imply Fe bulk velocities of $330_{-52}^{+93}$ km s$^{-1}$ and $483_{-78}^{+138}$ km s$^{-1}$, assuming an inclination of $i=30\pm5$ degrees, a column co-latitude of $\beta=61\pm5$ degrees \citep{1988MNRAS.231..597C}, where the dipole magnetic field line is tilted by $22\pm2$ degrees, and the binary radial-velocity semi-amplitude of WD at $K_1=107$ km s$^{-1}$ \citep{2002ASPC..261..167S}.
These velocities are lower than the expected post-shock velocity of $\sim 1{,}100$ km s$^{-1}$ for a WD mass of $M_{\rm WD}\sim 0.6\rm{-}0.7\,M_{\odot}$ \citep{1987MNRAS.226..209M,1995ApJ...455..260W,2022MNRAS.510.6110P}.  
This suggests the Fe lines arise farther down the accretion column (about $6\rm{-}16$\% of its height), where the post-shock plasma has cooled to $kT \sim 8\rm{-}12$ keV.
The bulk velocities measured with XRISM are smaller than the Chandra values reported by \citet{2007ApJ...658..525G}. 
Although this discrepancy could arise from different mass accretion rates between the observations, it does not affect the interpretation given above.
Furthermore, the velocity of Fe XXV is statistically smaller than that of Fe XXVI, allowing us to derive a velocity gradient (Figure \ref{fig:accretion_column_schematic}) within the accretion column of AM Herculis.
The distinct bulk Doppler maximum phases of Fe XXV ($0.10\pm0.01$) and Fe XXVI ($0.14\pm0.01$) suggest that the accretion column's inclination angle changes with height.

%-----------------------------------------------------------------------------------------------------------------
\subsection{Line width of Demodulated Fe spectra}
\label{section:results:demodulation}
%-----------------------------------------------------------------------------------------------------------------

%%===========================================================
%% Figure 6. (fig.4, de-modulation X-ray spectra)
%%===========================================================
\begin{figure*}[ht]
    \centering
    \includegraphics[width=0.95\linewidth]{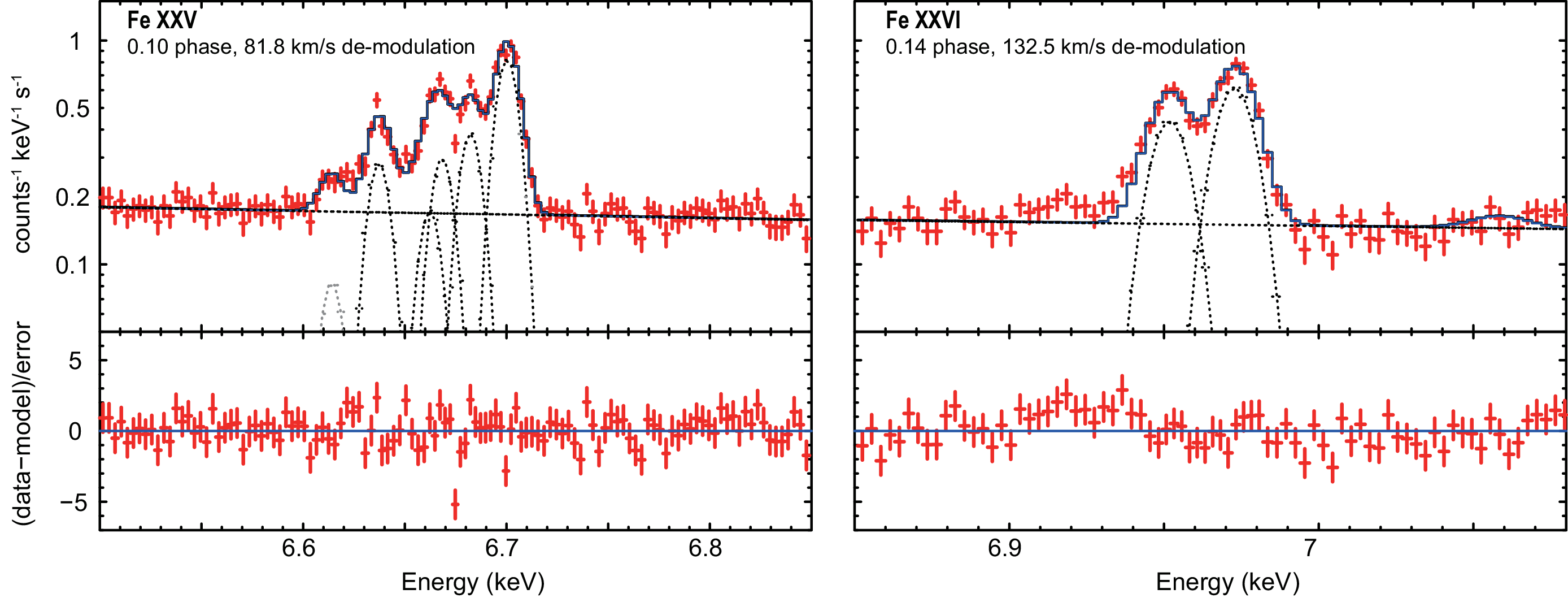}
    \caption{The demodulated X-ray spectra for Fe XXV and Fe XXVI lines are shown in the left and right panels, respectively, accompanied by the best-fit Gaussian models and the bremsstrahlung continuum. 
    The bulk Doppler shifts of X-ray energies across phases are demodulated using the best-fit modulation functions, represented by the cyan and magenta dashed lines in Figure \ref{fig:velocity2d_doppler} for Fe XXV and Fe XXVI, respectively.
    The lower panels illustrate the residuals relative to these models.
    }
    \label{fig:demodulate_spectra}
\end{figure*}
%%===========================================================

The bottom panel of Figure \ref{fig:velocity2d_doppler} (right) shows that the Doppler width of the Fe XXVI line is slightly broader than that of the Fe XXV line.
To examine this finding more clearly, we constructed Fe line spectra by demodulating the bulk Doppler shifts using the best-fit sinusoidal functions in Figure \ref{fig:velocity2d_doppler} (right). 
As a result, the demodulated X-ray spectra of the Fe XXV and Fe XXVI lines, displayed in Figure \ref{fig:demodulate_spectra}, exhibit Doppler widths of $5.23_{-0.15}^{+0.16}$ eV and $6.23_{-0.18}^{+0.19}$ eV, respectively, confirming that the Fe XXVI lines are substantially broader than the Fe XXV lines.
Because thermal Doppler broadening is the primary contributor to the line widths in the demodulated spectra, the XRISM data clearly show that Fe XXVI ions are located in a hotter region than Fe XXV ions, directly revealing the multi-temperature structure of the accretion column.
Previous studies \citep{1997ApJ...474..774F,2015A&A...578A..15L,2020A&A...642A.134S} estimated the plasma temperatures from line flux ratios between different ionization states of a given element and verified the multi-temperature structures using several atomic species.
In contrast, our XRISM observations show that the H-like and He-like ions of the same element originate from spatially distinct regions with different temperatures and velocities, implying that the previously reported absolute values will need to be revised.

%-----------------------------------------------------------------------------------------------------------------
\subsection{Anisotropy of Resonance lines}
\label{section:results:resonance}
%-----------------------------------------------------------------------------------------------------------------

%%===========================================================
%% Figure 7. (fig 5 top: EW vs phase, Enhance vs velocity) (EW Enhance vs Oscillator strength)
%%===========================================================
\begin{figure*}[ht]
    \centering
    \includegraphics[width=0.95\linewidth]{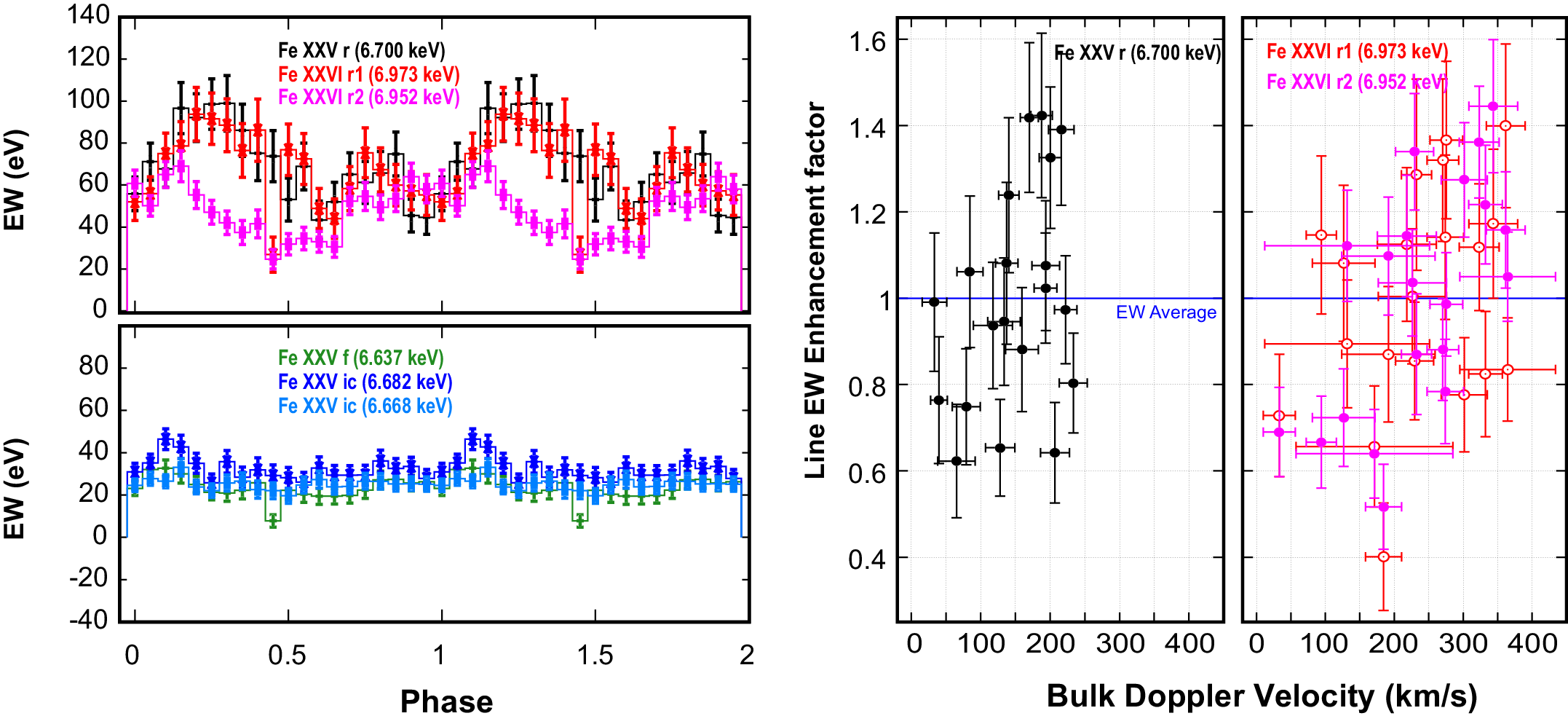}
    \caption{(Left) The upper and lower panels illustrate the equivalent widths of resonance lines and other lines across different phases, respectively. 
    The data definitions are indicated by labels in matching colors: r, r1, and r2 denote resonance lines, while f and ic denote the forbidden and intercombination lines, respectively.
    The errors are provided at a 1.0 $\sigma$ confidence level.
    (Right) The scatter plot shows the relationship between Doppler velocity and the enhancement of the line equivalent width relative to the mean values for the Fe XXV and Fe XXVI resonance lines, as shown in the left and right plots, respectively.
    }
    \label{fig:fe_ew_enhance}
\end{figure*}
%%===========================================================

The multi-temperature structure in the accretion columns of MCVs has a notable impact on the Fe line spectrum, as pointed out by \citet{1999PASJ...51...39T,2001MNRAS.328..112T}.
The accretion columns consist of optically thin thermal plasmas, but they are optically thick to resonance line scattering with large cross sections of $\sigma\sim 10^{-18}$ cm$^{2}$, corresponding to an optical depth of $\tau \sim 40$.
This situation is analogous to CIE plasmas in galaxy clusters, where resonance line photons traverse turbulent and bulk-flowing gas \citep{2016Natur.535..117H}.
In MCVs, the bulk flow is fast, at about $330\rm{-}480$ km s$^{-1}$ measured with XRISM, and thus, the resonance line energies are Doppler-shifted by amounts larger than the thermal width of $\sim 200$ km s$^{-1}$ (Figure \ref{fig:velocity2d_doppler} right).
As a result, as resonance line photons propagate within the accretion column, they may not be scattered due to the angular-dependent line energy shifts. 
Therefore, as illustrated in Figure \ref{fig:accretion_column_schematic}, this anisotropic opacity effect allows resonance line photons traveling vertically upward through the accretion column to escape without significant resonance scattering, while the plasma remains optically thick in the horizontal direction.
Therefore, the resonance Fe line emission should be strongly collimated along the vertical direction, as predicted by the first radiative transfer calculation conducted in \citet{2001MNRAS.328..112T}. 

Because line intensity can vary by the reflection component, the anisotropic resonance effect can be tested by checking whether the equivalent widths (EWs; line-to-continuum flux ratios) of resonance lines increase, which we expect in the pole-on phase, 
when the line of sight is nearly aligned with the vertical axis of the gas flow in the accretion column (hereafter we define the angle between them as the pole angle $\theta$, with a pole-on view corresponding to $\theta \sim 0$).
This test was first attempted observationally over 25 years ago with ASCA \citep{2001MNRAS.328..112T} and, more recently, with XMM-Newton \citep{2020A&A...642A.134S}, but the limited energy resolution introduced substantial systematic uncertainties, yielding inconclusive results.
The high-resolution XRISM spectra allow us to resolve resonance lines from neighboring lines and clearly detect their spin-phase modulations, as presented in Figures \ref{fig:spec_energy_phase_fe} and \ref{fig:demodulate_spectra}. 
We, therefore, measure the EWs of Fe resonance and other lines of XRISM over spin phase, as shown in Figure \ref{fig:fe_ew_enhance} (left), and find that only the resonance lines are significantly enhanced around phase $0.0\rm{-}0.2$.
To test whether this corresponds to the pole-on phase, we compare the EW enhancement factor (relative to the phase-averaged value) with the bulk Doppler velocity in Figure \ref{fig:fe_ew_enhance} (right).
The phase of large EWs roughly coincides with that of large Doppler velocity, i.e., the pole-on view. 
The fact that the largest increases in EWs occur at higher bulk Doppler velocities for all resonance lines is consistent with the resonance-anisotropy scenario in which reduced resonance opacity leads to enhanced emission, as illustrated in Figure \ref{fig:accretion_column_schematic}. 
This provides the first observational evidence of resonance anisotropy in MCV plasma.

%% 4. Discussions ================================================================================================
\section{Discussion}
\label{section:discussion}
%% ===============================================================================================================
%-----------------------------------------------------------------------------------------------------------------
\subsection{Resonance Scattering Anisotropy}
\label{section:discussion:resonance}
%-----------------------------------------------------------------------------------------------------------------

\begin{figure}[hbt]
    \centering
    \includegraphics[width=0.80\linewidth]{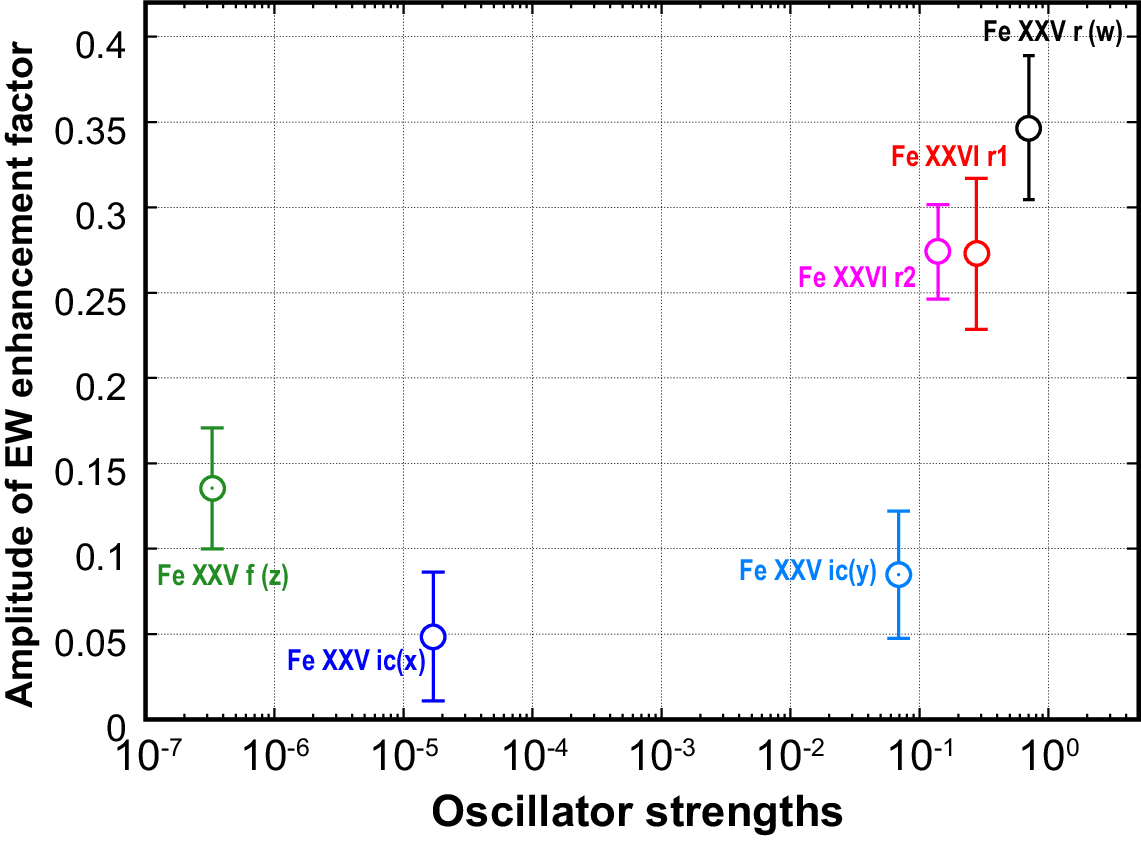}
    \caption{The illustration shows a scatter plot correlating the oscillator strengths of line transitions with the phase amplitude of the enhancement in equivalent widths. 
    It includes Fe XXV forbidden, intercombination 1 and 2, and resonance lines, alongside Fe XXVI resonance 1 and 2 lines, represented in green, cyan, blue, magenta, red, and black, respectively, with error bars indicating 1 $\sigma$ significance. 
    The oscillator strengths for Fe XXV lines are sourced from \citet{1999A&AS..135..347N}, and those for Fe XXVI are derived from \citet{1996ADNDT..64....1V}. 
    The vertical-axis values result from fitting the spin modulation of line equivalent widths in the upper-left panel of this figure to a sine function; they approximately correspond to the peak value of the right panel's vertical axis, minus 1.0.
    }
    \label{fig:fe_ew_enhance_oscillator}
\end{figure}

It is an interesting phenomenon that optically thin, hot plasma naturally and spontaneously exhibits anisotropic radiation in a cooling-flow environment.
The key point is that MCV plasmas exhibit "gray" opacity; specifically, the electron density is low enough to be optically thin for Compton scattering (with $\sigma\sim10^{-24}$~cm$^{-2}$), whereas the ion density is high enough to be optically thick for resonance scattering (with $\sigma\sim10^{-18}$~cm$^{-2}$). 
The probability of the resonance process -- encompassing photon absorption by a bound electron, excitation of the electron to a higher bound state, and subsequent photon re-emission --  is proportional to the oscillator strength of transitions from the ground state. 
Consequently, in the first-order approximation, a larger resonance anisotropy is expected for line transitions with higher oscillator strengths, thereby increasing opacity. 
Using the EW enhancement factor shown in Figure \ref{fig:fe_ew_enhance} (right), we determined their amplitudes and compared them with the oscillator strengths of the corresponding transitions, as displayed in Figure \ref{fig:fe_ew_enhance_oscillator}. 
The results show that the anisotropy of resonance photons positively correlates with the oscillator strengths of the transitions, providing further observational evidence for the anisotropic resonance scenario, first demonstrated by XRISM.
%
%%===========================================================
%% Figure 8. (fig 5 bottom: EW vs phase, Enhance vs velocity) (EW Enhance vs Oscillator strength)
%%===========================================================

%-----------------------------------------------------------------------------------------------------------------
\subsection{Modeling of Accretion Column Plasma properties}
\label{section:discussion:mcvspec_nustar}
%-----------------------------------------------------------------------------------------------------------------

%%===========================================================
%% Figure 9 (fig 10). Plasma structure
%%===========================================================
\begin{figure}[htb]
    \centering
    \includegraphics[width=0.5\linewidth]{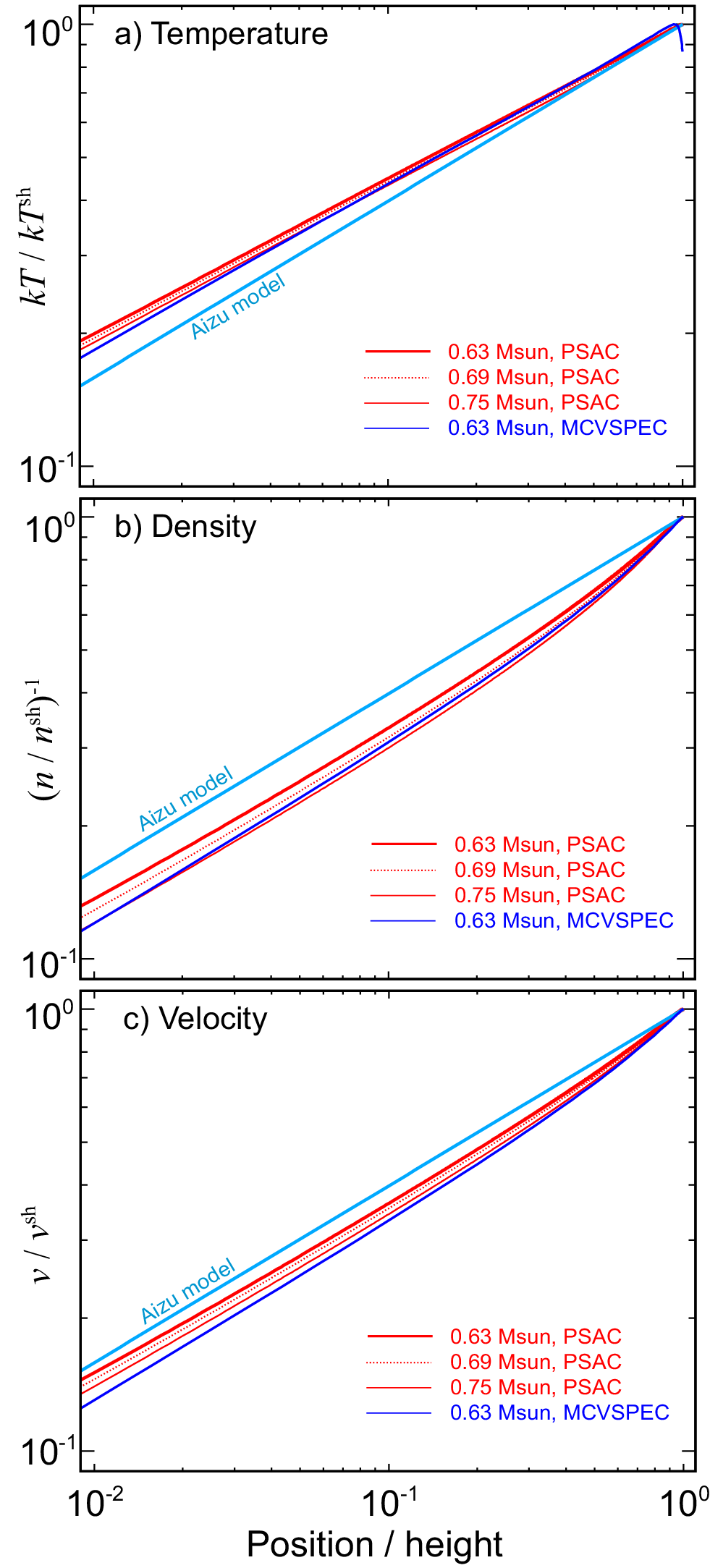}
    \caption{The position distributions of a) temperature, b) density, and c) velocity in the accretion column, as computed with the PSAC model \citep{2014MNRAS.441.3718H} and MCVSPEC \citep{Filor2025} for $M_{\rm WD}=0.63M_\odot$ and $B=13.6$ MG, are plotted in red and blue, respectively. For comparison, the Aizu model is also plotted in cyan.
    The x-axis represents the spatial coordinate scaled by the column height $h$.}
    \label{fig:methods:plasma_structure_psac_mcvspec}
\end{figure}

To examine whether the bulk velocities of Fe XXV and Fe XXVI measured with XRISM are compatible with the previously-inferred plasma structure of AM Her, we computed the plasma temperature, density, and velocity profiles, assuming a WD mass of $M_{\rm WD} = 0.63\,M_\odot$ \citep{Pala2020} and a magnetic field strength of $B = 13.6$\,MG,
from independent measurements \citep{1987MNRAS.226..209M, 1995ApJ...455..260W, 1998MNRAS.293..222C, 2007ApJ...658..525G, 2022MNRAS.510.6110P}.
%% PSAC
We first derived these profiles using the PSAC model \citep{2014MNRAS.441.3718H}, incorporating cyclotron cooling in addition to bremsstrahlung cooling and adopting a representative specific accretion rate of $\dot{m}$  = 1 g~cm$^{-2}$ s$^{-1}$. 
The resulting spatial profiles of the density $n$, bulk velocity $v$, and temperature $kT$, each normalized to the post-shock values $n^{\rm sh}$, $v^{\rm sh}$, and $kT^{\rm sh}$, are plotted as red curves in Figure \ref{fig:methods:plasma_structure_psac_mcvspec}.  
As expected, they differ from the simple analytic profiles of \citet{1973PThPh..49.1184A}, which are shown as cyan curves.  

%%===========================================================
%% Figure 10 (fig.9), NuSTAR + MCVSPEC
%%===========================================================
\begin{figure}[htb]
    \centering
    \includegraphics[angle=-90,width=0.90\linewidth]{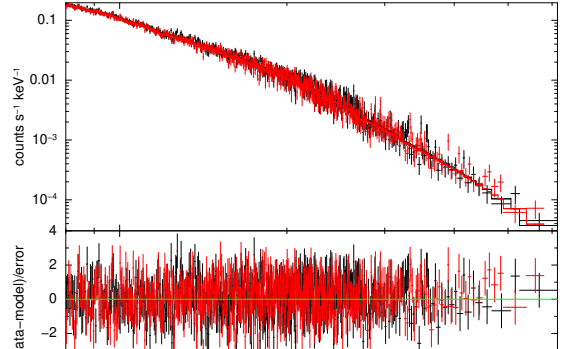}
    \caption{NuSTAR FPMA (black) and FPMB (red) spectra in the 8--60 keV band, fitted with the MCVSPEC model. 
    The NuSTAR spectra are rebinned to ensure at least 3-$\sigma$ signal-to-noise ratio in each bin. 
    No ISM absorption or Gaussian line components are required for these continuum-only X-ray spectra at higher energies. 
    We fixed $M_{\rm WD} = 0.63 M_\odot$ and $B = 13.6$~MG. 
    The MCVSPEC model fit yielded $\chi_\nu^2 = 0.97$ (1057 dof), $\dot{M} = 2.2\times10^{15}$~[g/s] and $f = 1.6\times10^{-4}$.}
    \label{fig:methods:nustar_mcvspec}
\end{figure}

%% MCVSPEC+NuSTAR
As a second approach, we analyzed the broadband X-ray spectra from the simultaneous NuSTAR observation using the MCVSPEC model \citep{Filor2025}. 
MCVSPEC is a new X-ray spectral model for mCVs that computes 1-D accretion column structures and outputs X-ray spectra in XSPEC for a given set of input parameters such as WD mass, magnetic field, fractional accretion column area, and abundances. Its application to X-ray spectra of another polar (EF Eri) and a detailed description of the model and fitting procedures are presented in \citet{Filor2025}. 
MCVSPEC self-consistently solves the coupled differential equations for the mass, momentum, and energy flows between the WD surface and the shock height.  
The model incorporates: (1) radiative cooling via thermal bremsstrahlung and electron cyclotron radiation, (2) two-temperature accretion flow with separate electron and ion profiles, (3) electron-ion energy exchange through Coulomb collisions, (4) gravitational acceleration within the accretion column, (5) the magnetic field geometry, and (6) X-ray reflection from the WD surface. 
%For each spatial grid in the accretion column at $z_i$, characterized by the plasma temperature $T(z_i)$, density $n(z_i)$, and bulk velocity $v(z_i)$, MCVSPEC computes the X-ray emissivity using the APEC and ATOMDB database, integrates it from $z=0$ (on the WD surface) to $z=h$ (at the shock height), and outputs an overall X-ray spectrum in XSPEC. 
%We accounted for the reflection of X-ray photons (emitted from the accretion column) off the WD surface by incorporating the {\tt reflect} model in XSPEC. For X-rays emitted from $z_i$ in the accretion column, we calculated the solid angle of the WD surface as seen from this height in the {\tt reflect} model. 
%This way, the X-ray reflection is considered for each layer and integrated over the accretion column later. 
%%
We fit the NuSTAR spectra in the 8--60 keV band where no X-ray absorption or atomic emission lines are present, as represented in Figure \ref{fig:methods:nustar_mcvspec}.
Since we fixed the WD mass at $M_{\rm WD}=0.63 M_\odot$ and the magnetic field strength at $B = 13.6$ MG in the MCVSPEC model, the only free parameters are the total mass accretion rate $\dot{M}$ [g/s] and the fractional accretion column area ($f$). 
The specific mass accretion rate, defined as $\dot{m} = \frac{\dot{M}}{4\pi R_{\rm WD}^2 f}$ where $R_{\rm WD}$ is the WD radius, determines the plasma temperature and density profiles in the accretion column sensitively \citep{Hayashi2014}. 
For a given $\dot{M}$ value (which controls the bolometric luminosity or flux normalization), varying $f$ (i.e., the column geometry) changes $\dot{m}$ and therefore the temperature profile, which alters both the overall X-ray flux and the spectral shape. Hence, given the known WD mass and magnetic field strength, we can uniquely determine $\dot{M}$ and $f$ for AM Herculis.  
Fitting the NuSTAR spectra with MCVSPEC yielded $\dot{M} = 2.2\times10^{15}$ [g/s] and $f = 1.6\times10^{-4}$, corresponding to $h/r = 1.2$ (Figure \ref{fig:methods:nustar_mcvspec}). 
%%%%%%%%%%
For these parameters, the specific mass accretion rate is 1.5 g\,cm$^{-2}$\,s$^{-1}$, validating the value of $\dot{m} = 1$ g\,cm$^{-2}$\,s$^{-1}$ assumed in the PSAC model. 
As a sanity check, we compare the bolometric luminosity ($L_{\rm bol}$) and the accretion luminosity $L_{\rm ac}\equiv GM_{\rm WD}\dot{M}/R_{\rm WD}$. 
First, we calculated the broadband X-ray luminosity (0.01--100 keV) by extrapolating the spectral models that fit the XRISM/Xtend and NuSTAR spectra over the 0.4--60 keV band. 
We also added the cyclotron luminosity $L_{\rm cyc} = 7\times10^{31}$~erg\,s$^{-1}$ estimated from \citet{Gansincke1995} scaled by a ratio of the X-ray fluxes of AM Hercules between 1990 and 2025, following the methodology applied to EF Eri \citet{Filor2025}). 
The estimated bolometric luminosity ($L_{\rm bol} = 1.9\times10^{32}$~erg\,s$^{-1}$) is consistent with $L_{\rm ac} = 2.2\times10^{32}$~erg\,s$^{-1}$. 

%%%
We finally derived the plasma temperature, density, and velocity at the shock height, as well as the geometrical parameters of the accretion column, which are summarized in Table \ref{tab:methods:plasma_param_mcvspec_rtsim}. 
The inferred shock temperature and velocities are consistent with those expected from the gravitational potential of a WD with $M_{\rm WD} = 0.63\,M_\odot$. 
The two approaches, using the PSAC and MCVSPEC models, yielded similar plasma temperature, density, and velocity profiles, as illustrated in Figure \ref{fig:methods:plasma_structure_psac_mcvspec}. 
Note that the discrepancy between $kT^{\rm sh}$ (= 24 keV) obtained from the MCVSPEC and NuSTAR data here and $kT_{\rm max}$ (= 37 keV; Section \ref{section:results:phase_averaged_overview} Table \ref{tab:methods:wide_band_fit}) is likely due to the simplified EM profile adopted in ‘bvcempow’, which assumes a power-law dependence of the differential EM on the plasma temperature $kT$.

%%%
Additionally, we examined the consistency with the XRISM observation of bulk Doppler modulations of Fe lines.
Given a shock velocity of $v^{\rm sh}=1{,}100$ km s$^{-1}$, the bulk Fe velocities probe regions near the heights $z$ of maximum emissivity for Fe XXV and Fe XXVI, at $z/h =0.06_{-0.02}^{+0.05}$ and $0.16_{-0.06}^{+0.13}$, respectively.  
The plasma model yields temperatures at these positions of $\sim 8.5_{-1.4}^{+2.1}$ keV for Fe XXV and $11.8_{-1.6}^{+2.7}$ keV for Fe XXVI.
These values correspond to the temperatures in regions where Fe ions are abundant and are roughly in agreement with the peak emissivity temperatures of 5.8 keV and 11.5 keV, respectively, from AtomDB \citep{2001ApJ...556L..91S}.
Therefore, the XRISM bulk-velocity measurements are in agreement with the plasma structure we derive for AM Her here.

%%===========================================================
%% Table 4, Simulation results
%%===========================================================
\begin{deluxetable*}{lcccccc}
\digitalasset
\tablewidth{0pt}
\tablecaption{Summary of the Plasma Parameters of the Accretion Column}
\tablehead{
    \colhead{Method}  &  \colhead{$kT^{\rm sh}$} & \colhead{$v^{\rm sh}$}  & \colhead{$n^{\rm sh}$}  &  \colhead{$h$} & \colhead{$r$}   & \colhead{$h/r$}\\
                      & (keV)           & (km s$^{-1}$)           & ($10^{15}$ cm$^{-3}$)   &  (km)          & (km)           &
%    \colhead{Parameter}             &  \colhead{Si XIV} & \colhead{S XVI}       & \colhead{Ca XX,XIV}   & \colhead{Fe XXV}      & \colhead{Fe XXVI}  & \colhead{Fe fluo}
}
\startdata
        MCVSPEC$^\dagger$    
                  & $24.0\pm0.1$   & $1,116\pm 2$  & $6.3\pm 1.2$       & $241 \pm 29$             & $223\pm 25$          &$1.08 \pm 0.01$\\
         \hline %-----------------------------------------------------------------------------         
        RT (Fe XXV r) $^\ddagger$ & \multicolumn{2}{c}{fixed above} 
                                        & $5.2_{-3.0}^{+45.0}$       & $280_{-260}^{+400}$     & $380_{-260}^{+210}$ & $0.7_{-0.5}^{+0.4}$\\
        RT (Fe XXVI) $^\ddagger$ & \multicolumn{2}{c}{fixed above} 
                                        & $6.7_{-3.4}^{+43.0}$       & $220_{-190}^{+230}$     & $340_{-210}^{+150}$ & $0.7_{-0.4}^{+1.0}$
\enddata
\tablecomments{$\dagger$ Best fit parameters obtained from fitting the MCVSPEC model to the NuSTAR spectra shown in Figure \ref{fig:methods:nustar_mcvspec}. The error bars are calculated from Markov chain Monte-Carlo simulations.\\
$\ddagger$ Radiative transfer simulation of resonance lines using the modified PSAC model. The ion names shown in brackets are the emission lines used in the simulation.}
\label{tab:methods:plasma_param_mcvspec_rtsim}
\end{deluxetable*}

%-----------------------------------------------------------------------------------------------------------------
\subsection{Resonance Radiative Transfer Simulations}
\label{section:discussion:rt_simulations}
%-----------------------------------------------------------------------------------------------------------------

%%===========================================================
%% Figure 11, Simulation results
%%===========================================================
\begin{figure}[htb]
    \centering
    \includegraphics[width=0.7\linewidth]{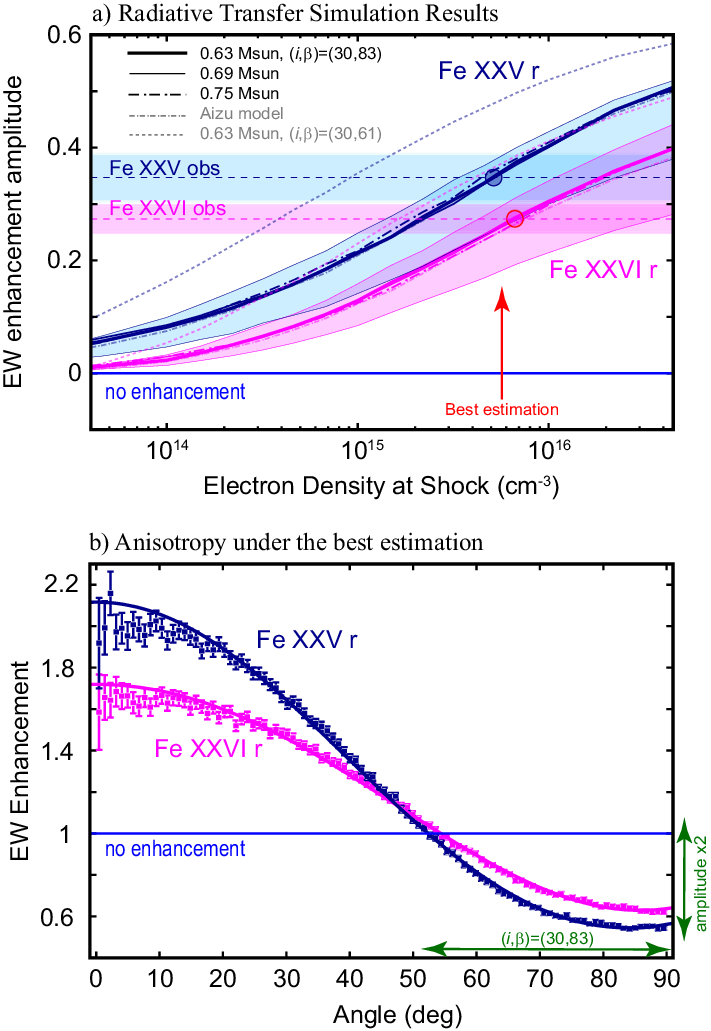}
    \caption{(top) The amplitudes of the EW enhancement factors, derived from RT simulations \citep{2001MNRAS.328..112T} for the AM Herculis observation (EM=$6.55\times10^{54}$ cm$^{-3}$), are displayed in relation to $n_{\rm e}^{\rm sh}$.
    Observed line amplitudes in Figure \ref{fig:fe_ew_enhance} are indicated by dashed, color-matched lines, and the conditions that best reproduce these values are highlighted with circles.
    Uncertainties due to errors in $i$, $\beta$, and in the observed EW\mbox{-}enhancement factor are represented using hatching.
    For comparison, the figure includes, in thin colors, the results assuming the Aizu model \citep{1973PThPh..49.1184A} and those for $0.63M_\odot$ that ignore the tilt of the magnetic field line.
    (bottom) The normalized EWs of the Fe XXV and Fe XXVI resonance lines, obtained from RT simulations, are shown in dark blue and magenta, respectively, as a function of the viewing angle of the accretion column under optimal conditions. 
    The viewing angle adopted in the top panel is marked by green arrows.
    }
    \label{fig:methods:simulation_results}
\end{figure}
%%===========================================================

%%% Configuration
The resonance anisotropy effect, revealed by XRISM, offers a new diagnostic of plasma density by combining the measured resonance line enhancement with the radiative\mbox{-}transfer (RT) Monte\mbox{-}Carlo simulations developed by \citet{2001MNRAS.328..112T}.
In order to numerically evaluate the resonance anisotropy, we performed these RT simulations, applying the temperature, density, and velocity structure from the PSAC model for $B = 13.6$\,MG (Figure \ref{fig:methods:plasma_structure_psac_mcvspec}) with incident Fe line emissivity by \citet{1985A&AS...62..197M}.
The accretion column is configured to reproduce the XRISM-observed emission measure of $EM=6.55\times10^{54}$ cm$^{-3}$ and NuSTAR predictions of $kT^{\rm sh}=24$ keV and $v^{\rm sh}=1{,}100$ km s$^{-1}$ from the MCVSPEC model.
Using this EM and the plasma structure derived in section \ref{section:discussion:mcvspec_nustar}, the post-shock electron density $n_{\rm e}^{\rm sh}$ becomes the sole remaining free parameter of the radiative transfer simulation of resonance lines.
We tried three cases of $M_{\rm WD}=0.63M_\odot, 0.69M_\odot$ and $0.75M_\odot$, regarding various independent measurements of $M_{\rm WD}$ in AM Herculis \citep{1987MNRAS.226..209M, 1995ApJ...455..260W, 1998MNRAS.293..222C, 2007ApJ...658..525G, 2022MNRAS.510.6110P}.
We conducted RT calculations spanning $n_{\rm e}^{\rm sh}$ from $10^{13}$ cm$^{-3}$ to $10^{16}$ cm$^{-3}$, extracting the enhancement amplitude of EWs through the viewing angle of the column between 53 and 113 degrees, with an error of $\pm 6$ degrees, which is estimated from $i$ and $\beta$ by \citet{1988MNRAS.231..597C} under the dipole magnetic field geometry.

%%% Results
Fundamentally, optically thick radiation, specifically the cyclotron cooling and resonance line emissions considered here, is sensitive to the column geometry and can distinguish between flatter "coin-like" and taller "cylinder-like" shapes.
The results of the RT simulations are presented in Figure \ref{fig:methods:simulation_results} (top) with the uncertainty in $n^{\rm sh}$ due to geometric errors in $i$ and $\beta$. 
We accumulated the Fe lines over pole angles $\theta$ between 53 and 90 degrees, following $i$ and $\beta$ values by \citet{1988MNRAS.231..597C}, assuming a dipole magnetic field line geometry.

%%%%%%%%%%%%%%%%%
As summarized in Table \ref{tab:methods:plasma_param_mcvspec_rtsim} and Figure \ref{fig:methods:simulation_results} (top), the best-fit $n^{\rm sh}$ values obtained from Fe XXV and Fe XXVI are consistent for all three WD-mass assumptions, yielding $n^{\rm sh} \approx (5\text{–}6)\times10^{15}$ cm$^{-3}$.
The corresponding best-fit column height and radius are $h \approx 200\text{–}300$ km and $r \approx 200\text{–}400$ km, giving a ratio of $h/r \sim 0.7\text{-}1.1$. 
In principle, optically thick radiation, such as resonance-line photons, is sensitive to the column geometry and can distinguish between coin-like and cylinder-like structures.
On the other hand, the MCVSPEC model fit to the broadband NuSTAR spectra allows us to determine the plasma temperature and density profiles robustly from optically-thin thermal bremsstrahlung spectra in the X-ray band. This is a complementary method for constraining the accretion column structure from the bulk velocity profile using phase-resolved XRISM spectroscopy data. 
As a result, we obtained a consistent set of accretion column parameters from the two approaches: the MCVSPEC modeling of the NuSTAR broadband spectra and the resonance-line radiative transfer analysis of the XRISM Fe line data.
This is the first demonstration that a combination of XRISM's Fe resonance line and NuSTAR's broadband X-ray continuum data, representing optically thick and thin plasma effects, enables the complete determination of all accretion column parameters.

%%% Discrepancy / Future prospect
Our results are roughly consistent with the $0.60M_{\odot}$ case by Chandra HETG in \citet{2007ApJ...658..525G}, except for the inferred values of the plasma scale of $h$ and $r$.
Such discrepancies can manifest as variations in EM at different observing epochs, and it may be necessary to adopt more realistic plasma conditions, including, for instance, more accurate cyclotron-cooling physics, multi-temperature bremsstrahlung components, and non-cylindrical accretion-column structures in a dipolar magnetic geometry, etc.
A comprehensive, more-realistic modeling of spin-resolved X-ray emission from MCV plasmas will be presented in a forthcoming paper.  

%%----------------------------
%%% Numerical results (electron density II)    
%%----------------------------
In CIE plasmas, an independent density diagnostic is provided by the ratio $R$ 
of the forbidden line $(f)$ to the intercombination lines $(x+y)$ \citep{1969MNRAS.145..241G,1981ApJ...249..821P,2001A&A...376.1113P}. 
In particular, the Fe XXV forbidden line, which arises from the radiative decay of the $^{3}S_{1}$ level to the ground state, can be suppressed by electron impacts that promote electrons from $^{3}S_{1}$ to the $^{3}P_{1}$ and $^{3}P_{2}$ levels, thereby enhancing the intercombination line emissions.
This technique has been applied to the Ne IX and O VII lines in MCVs \citep{2007ApJ...658..525G,2020A&A...642A.134S}.
However, a strong UV radiation field in MCVs provides an extra excitation channel from the $^{3}S_{1}$ to $^{3}P_{1}$ and $^{3}P_{2}$, artificially lowering $R$ and thus causing the density inferred from this ratio to be overestimated \citep{2001A&A...376.1113P,2001A&A...367..282N,2006ApJ...639..397I}. 
Accordingly, using the XRISM spectrum (Figure \ref{fig:demodulate_spectra}), we obtain $R=\frac{f}{(x+y)}\sim0.4$, which implies an overestimated density of $10^{17}$ cm$^{-3}$.
This inferred $R$ is about half of the $R \sim 0.7$ value derived from resonance-anisotropy diagnostics (implying an Fe density of $\sim$ a few $10^{15}$ cm$^{-3}$), suggesting that roughly 12\% of the total $(f+x+y)$ flux is influenced by UV photoexcitation.
Since the observed O VII ratio of $R\sim0.15$ corresponds to a much lower density of $\sim7\times10^{11}$ cm$^{-3}$ \citep{2020A&A...642A.134S} than the Fe-based value here, a more comprehensive analysis is required either to refine the $R$-ratio density diagnostic or to explore alternative accretion scenarios for MCVs.

%-----------------------------------------------------------------------------------------------------------------
\subsection{Origin of Fe Fluorescent lines}
\label{section:discussion:fluorescent}
%-----------------------------------------------------------------------------------------------------------------

%%% Fluorescent  (From Result section)
The XRISM spectrum indicates a large Doppler width of the Fe fluorescent line exceeding 400 km s$^{-1}$ (Table \ref{tab:methods:line_fit}).
In principle, Fluorescent lines are produced in cold material and are typically interpreted as reprocessed X-rays from the WD surface, which is irradiated by intense X-rays from the accretion column.
However, the WD surface itself cannot reach such a high velocity; it is only about 5 km s$^{-1}$ at a WD radius of $8\times10^8$ cm, given the AM Herculis spin period of 11139.3 s \citep{2020A&A...642A.134S}.
Thus, the measured velocity implies a non-negligible contribution from fluorescence in a cold accretion flow with a velocity of $\sim 400$ km s$^{-1}$.
This interpretation is supported by the phase$\rm{-}$energy distribution of the Fe fluorescence (Figure \ref{fig:spec_energy_phase_fe} lower), which appears to show a spin-modulated component and an unmodulated component from the WD surface.
The observed line width of $\sim$400 km s$^{-1}$ lies between the velocity of the WD surface and the free-fall velocity just before the shock, which reaches about 4,700 km s$^{-1}$ for $M_{\rm WD}\sim0.7M_\odot$. This suggests that an additional fluorescent component may originate in the gas flow located well above the shock region.
Very recently, X-ray Doppler tomography of the Fe K line has been successfully performed for the low-mass X-ray binary 4U~1822–371 using XRISM high-resolution spectroscopy \citep{2026arXiv260304679S}. This new technique will constrain the emission region of the fluorescent line in the AM Herculis system. The detailed results will be presented in a future publication.

%% 4. Conclusion =================================================================================================
\section{Conclusion}
\label{section:conclusion}
%% ===============================================================================================================

We performed the X-ray observation of AM Herculis with XRISM and NuSTAR. We highlighted our main results in the following points.

\begin{enumerate}
    %%%% Item (1) lines in averege spectrum %%%%%%%%%
    \item 
    In the XRISM high-resolution spectrum of AM Herculis, we have clearly resolved all satellite lines associated with highly ionized Fe from MCV plasma for the first time (Section \ref{section:results:phase_averaged_lines}), as well as lines from lighter elements: Si, S, and Ca.
    The light elements show relatively narrow line widths of 2–3 eV, consistent with thermal broadening alone, whereas the Fe lines exhibit larger widths of 6–7 eV, indicating an additional Doppler broadening effect arising from bulk gas motions.
    
    %%%% Item (2) spin-Doppler modulation %%%%%%%%%
    \item 
    We clearly detect spin-phase-dependent Doppler modulations in the Fe lines (Section \ref{section:results:phase_resolved}) with XRISM. 
    The Fe XXV and Fe XXVI lines show modulation semi-amplitudes of $81.8\pm6$ km s$^{-1}$ and $132.5\pm9$ km s$^{-1}$, and mean velocities of $143.6\pm6$ km s$^{-1}$ and $225.6\pm8$ km s$^{-1}$, respectively, implying Fe bulk velocities of $330_{-52}^{+93}$ km s$^{-1}$ and $483_{-78}^{+138}$ km s$^{-1}$ for a system geometry by \citet{1988MNRAS.231..597C}.
    The difference between the two ions points to bulk velocity gradients along the accretion column, while their absolute values, well below the expected post-shock velocity of $\sim 1{,}100$ km s$^{-1}$ for $M_{\rm WD}\sim 0.6\rm{-}0.7\,M_{\odot}$ \citep{1987MNRAS.226..209M,1995ApJ...455..260W,2022MNRAS.510.6110P}, indicate that the Fe-emitting region with $kT \sim 8\rm{-}12$ keV lies lower in the column at about $6\rm{-}16$\% of its height.
    After removing the modulation, the Fe XXV and Fe XXVI lines have Doppler widths of $5.23_{-0.15}^{+0.16}$ eV and $6.23_{-0.18}^{+0.19}$ eV, respectively, demonstrating that Fe XXVI is significantly broader and arises from a hotter region than Fe XXV (Section \ref{section:results:demodulation}). 
    These results constrain the dynamics of the hot accretion plasma and directly reveal the multi-temperature structure of the accretion column.
    
    %%%% Item (3) resonance enhancement %%%%%%%%%
    \item 
    Phase-resolved XRISM spectra clearly demonstrate the anisotropy of Fe resonance lines (Section \ref{section:results:resonance} and Section \ref{section:discussion:resonance}).
    The EW of resonance lines from Fe XXV and Fe XXVI are enhanced from the phase\mbox{-}average by factors of 1.30 -- 1.35, and these enhancements show a positive dependence on the oscillator strengths of the responding transitions for Fe XXVI resonance, Fe XXV resonance, intercombination, and forbidden lines.
    Such anisotropy was predicted more than 25 years ago \citep{1999PASJ...51...39T,2001MNRAS.328..112T} and our results present the first observational evidence of this prediction.
    
    %%%% Item (4) plasma structure %%%%%%%%%
    \item 
    We derive a unique, self-consistent solution for the accretion column structure that simultaneously explains both the XRISM and NuSTAR data, combining accretion plasma models (PSAC and MCVSPEC; \citealt{2014MNRAS.441.3718H,Filor2025}) with radiative transfer simulations of resonance photons \citep{2001MNRAS.328..112T} (Section \ref{section:discussion:mcvspec_nustar} and Section \ref{section:discussion:rt_simulations}).  
    Adopting $M_{\rm WD}=0.63M_\odot$, $B=13.6$ MG \citep{Pala2020}, and a specific accretion rate of $\dot{m}=1$ g~cm$^{-2}$ s$^{-1}$ with EM=$6.55\times10^{54}$ cm$^{-3}$, the broadband NuSTAR X-ray spectrum characterized with the MCVSPEC model yields $kT^{\rm sh}=24.0\pm0.1$ keV and $v^{\rm sh}=1{,}116\pm2$ km s$^{-1}$.  
    Using the plasma structure predicted by the PSAC model under these conditions, the RT simulations reproduce the observed line enhancements in the XRISM high-resolution spectra for densities of $n^{\rm sh} \approx (5\text{–}6)\times10^{15}$ cm$^{-3}$, heights of $h \approx 200\text{–}300$ km, and radii of $r \approx 200\text{–}400$ km, which correspond to a geometric aspect ratio of $h/r \sim 0.7\text{-}1.1$.

    %%%% Item (5) plasma diagnostics %%%%%%%%%
    \item The resonance anisotropy provides new plasma density diagnostic, independent of methods based on the $R$ ratio (Section \ref{section:discussion:rt_simulations}).  
    By comparing the results from these two diagnostic approaches, we conclude that approximately 12\% of the total $(f+x+y)$ line flux is affected by UV photoexcitation, which thus limits the reliability of $R$-ratio-based density diagnostics in MCVs.

    %%%% Item (6) fluorescent %%%%%%%%%
    \item 
    In addition to investigating plasma dynamics with highly ionized Fe lines above, we identified a broad fluorescent Fe-K line whose width exceeds the Doppler broadening expected from the WD surface motion, indicating a non-negligible contribution from a cold accretion flow with a velocity of $\sim 400$ km s$^{-1}$ (Section \ref{section:discussion:fluorescent}).

\end{enumerate}

Our detailed X-ray spectral study of AM Herculis yields the first direct structure of the shock-heated, magnetically confined plasma of MCVs that leads to anisotropic resonance-line transfer. 
This resonance effect is also expected for lighter elements originating from cooler, lower regions of the column and implies that spin-phase-dependent atomic X-ray line spectra must be incorporated in future astrophysical analyzes.
Although this anisotropy occurs only in resonant scattering and has little impact on the overall luminosity, revisiting the line spectra of MCVs over their spin phases is crucial for a wide range of astrophysical objectives, including investigating the binary evolution of MCVs driven by mass transfer in potential type Ia supernova progenitors and tackling the long-standing puzzle of the origin of the GC and GRXE emissions.  
Moreover, the structure of strongly shock-heated, magnetically confined accretion plasmas is of interest not only for its fundamental role in spontaneous collimation phenomena but also for its relevance to laboratory plasma physics.

%% Acknowledgement  ==============================================================================================
\begin{acknowledgments}
The authors thank all members of the XRISM team, especially for their continuous efforts in operating the spacecraft and maintaining the onboard instruments.
This work was supported by the JSPS Core-to-Core Program (grant number: JPJSCCA20220002) and the Japan Society for the Promotion of Science Grants-in-Aid for Scientific Research (KAKENHI) Grant Number JP20K04009 (YT), JP26H01396 (YT) and JP24K00677(MN).
KM, GB and SW are supported by NASA's XRISM AO-1 grant (80NSSC25K7539) and ADAP program (80NSSC23K0971). 
The authors gratefully acknowledge the  supportive and constructive comments and suggestions provided by the anonymous referee.
\end{acknowledgments}
%% ===============================================================================================================

%% Contribution  =============================================================================================
\begin{contribution}
YT was responsible for writing and submitting the manuscript, and, as PI of the XRISM observation of AM Herculis, also originated the research concept and conducted the observations, data analysis, radiative transfer simulations, and interpretation of the results.  
KM served as co-PI for XRISM and PI for the NuSTAR observations, conceived the research concept, performed the observations, and edited the manuscript.  
TH and GB performed the analyses and computations using plasma models, and AM independently cross-checked the XRISM data analysis.  
MI, AS, MK, TI, and MT formulated the initial observational plan, contributed to the development of analysis methods, and revised the manuscript.  
MN and DB likewise contributed to the analysis methodology and participated in manuscript revision.  
SB, CH, GR, AR, and SW provided feedback and comments on the manuscript.
\end{contribution}
%% ===============================================================================================================

%% Facilities ====================================================================================================
\facilities{XRISM(Resolve), 
            NuSTAR(FPMA,FPMB), 
            AAVSO}

\software{XSPEC \citep{1996ASPC..101...17A},
          PSAC \citep{2014MNRAS.441.3718H},  
          MCVSPEC \citep{Filor2025},
          MCV radiative transfer simulation \citep{2001MNRAS.328..112T}
          }
%% ===============================================================================================================

%% Appendix  =============================================================================================
%\appendix

%% Bibliography =============================================================================================
\bibliography{xrism_amher}
\bibliographystyle{aasjournalv7}

\end{document}